\documentclass[twocolumn,biblatex]{customtemplate}

\usepackage[english]{babel}
\usepackage[T1]{fontenc}
\usepackage{subcaption}
\usepackage{graphicx}
\usepackage{combelow}
\usepackage{mathtools}
\usepackage{amsfonts, amssymb}
\usepackage{sidecap}
\usepackage{booktabs}
\usepackage{cleveref}

\newcommand{\Unif}[2]{\ensuremath{\mathcal{U}_{\left[#1,#2\right]}}}
\newcommand{\lik}{\mathcal{L}_D}

\newcommand{\moment}[1]{\mathbb{E}\left[\mathcal{X}^{#1}\right]}
\renewcommand{\d}{\mathrm{d}}
\newcommand{\front}[1]{\ensuremath{\bar{#1}}}
\newcommand{\back}[1]{\ensuremath{\tilde{#1}}}
\renewcommand{\vec}[1]{\boldsymbol{#1}}

\usepackage{lipsum}

\usepackage{hyperref}

\title{\Large \bf{A nonlinear theory for chemotactic fronts of mixed populations}}% Force line breaks with \\
\author{\centering \normalsize Giulia L. Celora$^{1,2,*}$\;,  Marjorie Watts$^{1}$\;, Carles Falc\'o$^{1}$\;,  Mohit P. Dalwadi$^{1}$}

\date{}% It is always \today, today,
\begin{document}
\hypersetup{
  colorlinks=true,
  allcolors=black
}
\twocolumn[\maketitle\par\vspace{-1.3cm}
\begin{center}
{$^{1}$Mathematical Institute, University of Oxford, OX2 6GG Oxford, United Kingdom}

{$^{2}$School of Mathematics, University of Bristol, Fry Building, BS8 1UG Bristol, United Kingdom}

{$^*$Correspondence: giulia.celora@maths.ox.ac.uk}
\par\vspace{5ex}
\begin{minipage}{16cm} 
Collective migration of heterogeneous cell populations is central to many biological and physiological processes, including development and immune response. Recent experimental and theoretical advances have shown how asymmetric interactions with self-generated chemical gradients shape the spatial distribution of distinct cell types within migrating collectives. However, the principles governing robust spatial organisation of heterogeneous cell populations remain poorly understood. Here, we use asymptotic analysis to systematically derive a nonlinear analytical theory for heterogeneous cell collectives guided by self-generated chemotaxis. Our theory disentangles how heterogeneity in cell diffusivity, chemoattractant consumption, and chemotactic sensitivity shape the density profiles of migrating heterogeneous collectives, revealing four distinct dynamical behaviours that together capture all possible regimes. We calibrate our framework to experimental data on the co-migration of dendritic and T cells. We predict that this system operates in a parameter regime that balances intercellular mixing with T-cell localisation at the leading front of the migrating collective. Our theory reveals that this behaviour is enabled by intermediate long-range chemoattractant signalling generated through strong chemoattractant consumption by dendritic cells. Overall, our framework provides general principles for understanding how non-reciprocal chemical interactions shape robust collective migration in heterogeneous cell populations.

\end{minipage}
\end{center}

\noindent \small{\emph{Keywords:} self-generated chemotaxis $|$ asymptotic analysis $|$ cellular heterogeneity $|$ nonlinear waves}
\par\vspace{5ex}
]

\addtocontents{toc}{\protect\setcounter{tocdepth}{0}}

%\tableofcontents

\section*{Introduction}
Long-range coordinated migration of cell collectives is a fundamental process across biological systems, from development and wound healing to immune response and cancer metastasis \cite{cheung_collective_2025,wu_collective_2025,stehbens_perspectives_2024}. In many contexts, this coordination is mediated by self-generated chemotaxis, whereby cells both shape and respond to chemoattractant gradients that they create through consumption \cite{alanko_ccr7_2023,cremer_chemotaxis_2019,dona_directional_2013,ford_pattern_2025,kiran_macrophage_2025,malet-engra_collective_2015,tweedy_seeing_2020,Ucar2025}. This mechanism has been shown to efficiently drive long-range migration, often outperforming externally imposed gradients~\cite{tweedy_self-generated_2020}.

In \emph{in vivo} settings, collective migration typically involves multiple cell types with distinct abilities to sense, respond to, and modify chemical gradients~\cite{mclennan_multiscale_2012,wu_collective_2025}. Such heterogeneity is observed across a wide range of systems, from bacterial populations~\cite{fu_spatial_2018} to leader–follower dynamics in neural crest migration during chicken embryo development~\cite{mclennan_multiscale_2012}, the zebrafish lateral line~\cite{wu_collective_2025}, and immune cell interactions~\cite{Ucar2025}. Despite this, theoretical and experimental studies of self-generated chemotaxis have primarily focused on homogeneous populations. As a result, we are just starting to understand how cellular heterogeneity shapes collective migration.
Theoretically, there has been a growing interest in the role of cellular heterogeneity in chemotaxis-based mechanisms, such as patterning via aggregation-diffusion and long-range migration. Building on the seminal work of Keller–Segel~\cite{keller_traveling_1971}, reaction–advection–diffusion frameworks have been widely used to study trait-structured chemotaxis, where heterogeneity commonly arises from variability in cells’ chemotactic responses~\cite{fu_spatial_2018,ridgway_motility-induced_2023,Salek2019TmazeChemotaxis,mattingly_collective_2022,gude_bacterial_2020,bai_spatial_2021,vo_nongenetic_PNAS_vo} and cells' interactions with the chemical environment~\cite{macfarlane_impact_2022,lorenzi_phenotype_2025,freingruber_trait-structured_2025,Ucar2025}. These phenomenological models have explored both discrete~\cite{macfarlane_impact_2022,fu_spatial_2018,Ucar2025} and continuous forms of cellular heterogeneity~\cite{lorenzi_phenotype_2025,freingruber_trait-structured_2025,mattingly_collective_2022}, as well as the role of natural selection and/or local environment in shaping cells' trait distribution~\cite{lorenzi_phenotype_2025,macfarlane_impact_2022,ridgway_motility-induced_2023,mattingly_collective_2022}. These approaches have provided important insights into how heterogeneous cell populations generate spatial patterns through aggregation~\cite{lorenzi_phenotype_2025,macfarlane_impact_2022,freingruber_trait-structured_2025,ridgway_motility-induced_2023} and organise within migrating fronts~\cite{fu_spatial_2018,Ucar2025,mattingly_collective_2022}. However, only a limited number of studies have quantitatively validated these theoretical predictions against experimental data~\cite{fu_spatial_2018}, and only recently, in the context of eukaryotic cell migration~\cite{Ucar2025}. We therefore still lack a comprehensive understanding of chemotaxis-guided co-migration of heterogeneous eukaryotic cell populations, and the physical principles that shape trait-specific spatial organisation of cells within the collective. %, 

Here, we address this gap by building on the recent work of~\parencite{Ucar2025}, which investigates how self-generated gradients can serve as a minimal mechanism for the interaction and migration of heterogeneous cell populations. In particular, Ucar et al.~focus on the co-migration of two cell types that are (asymmetrically) coupled through a diffusible chemoattractant. Specifically, the asymmetry arises from an assumed trade-off between the ability of cells to consume and sense the external chemoattractant. As a result, one cell type (the consumers) specialises in chemoattractant consumption and exhibits weak chemotaxis, while the other (the sensors) does not consume the chemoattractant but has greater chemotactic sensitivity. The resulting sensor/consumer (SC) model predicts that sharp self-generated gradients allow for cell populations to migrate as travelling waves, without the need for mechanical cell–cell interactions. Further numerical investigation of the predicted spatial structure of travelling-wave fronts reveals optimal co-migration regimes that allow the two distinct populations to migrate robustly as a spatially-structured collective. These theoretical predictions were compared with data from an \emph{in vitro} microfluidic assay with mixtures of dendritic and T cells. However, their model calibration relies primarily on a linearisation of the SC model, which enables computation of the front speed and leading-edge structure, but constrains only a subset of model parameters.

In this paper, we present a combined calibration and analytical study of the SC model proposed by~\cite{Ucar2025}, showing that the spatial structure of the migrating front contains additional information about inter-population coupling and gradient formation. In particular, we demonstrate that nonlinear features of the migrating front, neglected in the analysis of~\cite{Ucar2025}, allow for full quantitative characterization of the co-migration of sensor-consumer systems. Through quantitative calibration of the model to experimental data for the spatial cell densities, we confidently identify the parameter regimes that describe the dendritic and T cells' co-migration. This analysis reveals that the system falls within a distinguished limit of the model, where sharp chemoattractant gradients are balanced by weak chemotactic strength. Leveraging asymptotic techniques, we characterise the full phase diagram of the SC model and systematically derive asymptotic analytical expressions that directly link model parameters to the spatial organisation of heterogeneous cell populations within the migrating front. With this, we provide a complete investigation of robust co-migration strategies across parameter regimes. Our nonlinear theory offers a comprehensive characterisation of the SC system, and highlights the physical principles that guide sustained and robust co-migration of heterogeneous cell populations. 

\section{Mathematical Model}
We investigate the model proposed in~\cite{Ucar2025}, which describes the coordinated dynamics of a binary mixture of chemotacting cell populations and successfully captures the spatiotemporal evolution of dendritic and T-cell densities, respectively denoted by \(\rho=\rho(\vec{x},t)\) and \(\eta=\eta(\vec{x},t)\), during co-migration. In this framework, the long-range collective migration of the two immune cell subpopulations emerges from three key mechanisms: (i) random motion of the two cell populations; (ii) directed motion of both cell populations in response to an external chemical signal \(a=a(\vec{x},t)\); (iii) dendritic cells (the consumers) degrade the chemical. For simplicity, the model neglects consumption of the chemoattractant by the T-cell. This approximation is consistent with experiments showing that T cells migrating in isolation fail to sustain the coherent wave-like migration observed during co-migration, suggesting that their contribution to reshaping the chemoattractant field is small compared with that of dendritic cells~\cite{Ucar2025}. Together, these assumptions lead to a modified version of the standard Keller--Segel model,
\begin{subequations}
    \begin{align}
      \frac{\partial a}{\partial t} &=\tilde{D}_a\nabla^2 a-\tilde{r}_\rho a\rho,\label{eq:time-dependent-a}\\
        \frac{\partial\rho}{\partial t}&=\nabla\cdot\left(\tilde{D}_\rho\nabla\rho-\tilde{\chi}_\rho\, \rho\ \nabla \ln (a+a_m)\right),\\
         \frac{\partial\eta}{\partial t}&=\nabla\cdot\left(\tilde{D}_\eta\nabla\eta-\tilde{\chi}_\eta\, \eta\ \nabla \ln (a+a_m)\right),
    \end{align}\label{eq:time-dependent problem_a}%
\end{subequations}
where$\tilde{D}_{\rho,\eta}$ and $\chi_{\rho,\eta}$ indicate the diffusion coefficients and the chemotactic coefficients of the two cell populations, while $\tilde{D}_a$ and $\tilde{r}_\rho$ denote the chemoattractant diffusion coefficient and the rate at which the chemoattractant is consumed by the consumer population. Eq~\eqref{eq:time-dependent problem_a} is supplemented with appropriate initial and boundary conditions (Section A.1, Supplementary Information). 

As discussed in~\cite{Ucar2025}, when considering Eq.~\eqref{eq:time-dependent problem_a} on a semi-infinite 1D line ($x\in\mathbb{R}^+$), stable travelling-wave solutions mimicking long-range collective migration can be obtained in the limit of a small sensing threshold $a_m\to0^+$, when the concentration of consumers $\rho^\dagger$ at the back of the wave is maintained constant -- for example via imposing a constant influx of consumer cells (Section A and Figure S1, Supplementary Information). These two assumptions together ensure that cells at the rear of the group do not lag behind the front by maintaining a constant, nonzero chemotactic signal \(S=\partial_x a/a\) at the back of the wave. As a result, eventually, the travelling group migrates with a constant velocity $v$, which is set by both the properties of the consumer cells and the chemoattractant via the expression~\cite{Ucar2025}
\begin{equation}
v=\sqrt{\dfrac{\rho^{\dagger}\tilde{r}_\rho\tilde{\chi}^2_\rho}{\tilde{D}_a+\tilde{\chi}_\rho}}\label{eq:velocity},
\end{equation}
where $\rho^{\dagger}>0$ is the asymptotic value of the consumer population at the back of the wave; the latter can be expressed in terms of cell influx~\cite{marjorie2026}. Although the analytical form for migration speed is derived by~\parencite{Ucar2025} via linearisation of~\eqref{eq:time-dependent problem_a} at the back of the wave front, previous work investigates the spatial structure of the migrating front predicted by~\eqref{eq:time-dependent problem_a} only numerically. Here, we investigate travelling-wave solutions to~\eqref{eq:time-dependent problem_a} for a moving 1D front on $x\in\mathbb{R}$ more generally and fully characterise the spatial organisation of the two populations $\rho$ and $\eta$ within the propagating front defined by the travelling wave. 
Under the assumption that $a_m=0$, we can obtain a closed-form equation for the chemotactic signal $S=\partial_x\ln a$ from~\eqref{eq:time-dependent-a}. This is equivalent to computing the inverse \emph{Cole--Hopf transform} for the chemical density field -- an approach commonly applied to study quasi-linear transport problems~\cite{bok_diff_equations}. Using this transformation, and considering a 1D domain, we can rewrite~\eqref{eq:time-dependent problem_a} as
\begin{subequations}
        \begin{align}
         \frac{\partial S}{\partial t} &=\tilde{D}_a\partial_{x}(\partial_xS+S^2)-\tilde{r}_\rho \partial_x\rho,&\label{eq:burger_chemoattractant}\\
           \frac{\partial\varphi}{\partial t}&=\partial_{x}(\tilde{D}_\varphi\partial_x\varphi-\tilde{\chi}_\varphi S\varphi),&\quad \varphi\in\left\{\rho,\eta\right\}.
    \end{align}\label{eq:time-dependent problem}%
\end{subequations}
Eq.~(\ref{eq:burger_chemoattractant}) is equivalent to the viscous Burger equation with an additional source term to account for the role of consumption in shaping the chemotactic drift $S$. This formulation makes explicit that the shape of the chemotactic signal $S$ is affected by two different transport mechanisms that evolve on different timescales -- namely, diffusion and a quasi-linear advection.

\begin{figure*}[th!]
\centering
\includegraphics[width=\textwidth]{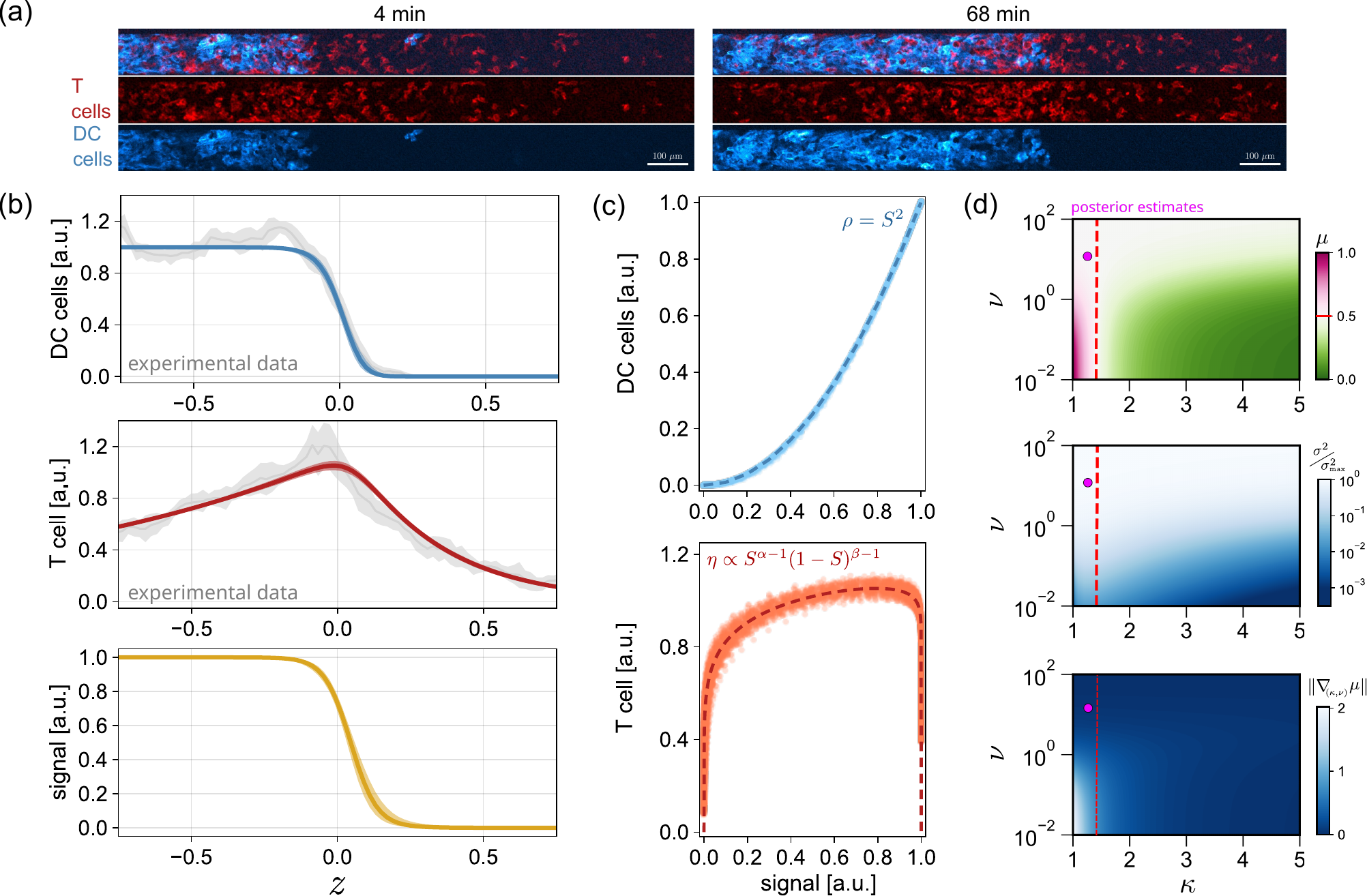}
\caption{\textbf{Predictions for the coupled co-migration of T and dendritic cells.} (a) Microscopy images from the microfluidic channel experiments from~\cite{Ucar2025} with labelled DCs (blue) and co-migrating T cells (red) at two time points: (Left) $4$ min and (Right) $68$ min. (Obtained from~\cite{Ucar2025}, licensed under CC BY 4.0). (b) Posterior prediction of the model~\eqref{eq:travelling_wave_population} and comparison with experimental data (gray curve and shaded area) for the density of dendritic cells $\rho$ (blue), and the density of the T cells $\eta$ (red). For completeness, we also illustrate the predicted chemotactic signal $S$ (gold), even if this is not measured experimentally. Details on the calibration are given in Section B, Supplementary Information. (c) Posterior predictions for the dendritic cells (top) and T cells (bottom) as a function of the chemotactic signal $S$ as predicted by the calibrated model (dots). The dashed line indicates the profile predicted by the asymptotic theory taking $\chi_\rho\to 0$ (Section C, Supplementary Information). (d) Colormaps illustrating the theoretical predictions for the mixing metrics $\mu$, $\sigma$ as a function of $(\kappa,\nu)$ in the limit $\chi_\rho\to 0$. The red dashed line indicates the isocline $\mu(\kappa,\nu)\equiv0.5$, while the magenta dots indicate the estimated value of the model parameters obtained by fitting~\eqref{eq:travelling_wave_population} to the experimental data from~\cite{Ucar2025} (see panel (a) and Figure S2, Supplementary Information).}\label{fig:fitting}
\end{figure*}

\subsection*{Travelling wave problem}
In 1D and denoting the spatial dimension via $x\in\mathbb{R}$, we look for travelling wave solutions to~\eqref{eq:time-dependent problem} of the form
\begin{equation}
    \varphi(x,t)=\varphi(z), \quad \tilde{z}=x-vt, \quad \varphi\in\left\{\rho,\eta,S\right\},\label{eq:transformation TW}
\end{equation}
where $v>0$ is the velocity of the travelling wave~\eqref{eq:velocity}. Substituting~\eqref{eq:transformation TW} into~\eqref{eq:time-dependent problem}, integrating and non-dimensionalising the system (as detailed in Section A, Supplementary Information) reduces the problem to the following set of coupled nonlinear ordinary differential equations:
\begin{subequations}
    \begin{align}
       D_\rho\rho'&=\left(S- 1\right)\rho,\label{eq:TW eq rho}\\
       \quad D_\eta\eta'&=(\kappa S-1)\eta,\label{eq:TW eq eta}\\ 
       \chi_\rho S'&=(1+\chi_\rho)\rho-S(S+\chi_\rho),\label{eq:TW eq S}
    \end{align}\label{eq:travelling_wave_population}%
    where $\varphi':=D_av^{-1} \ \mathrm{d}\varphi/\mathrm{d}z$ indicates the rescaled derivative of the density profile with respect to the travelling-wave coordinate $z=\tilde{z}v/\tilde{D}_a$. Eqs.~\eqref{eq:TW eq rho}-\eqref{eq:TW eq S} have a total of four non-dimensional parameters $\chi_\rho=\tilde{\chi}_\rho/\tilde{D}_a$, $D_{\rho,\eta}=\tilde{D}_{\rho,\eta}/\tilde{D}_a$ and $\kappa=\tilde{\chi}_\eta/\tilde{\chi}_\rho>1$, which characterise the ratio of the consumer chemotactic sensitivity and cell diffusivities to the chemoattractant diffusivity, and the relative chemotactic sensitivity of the two cell populations. Given the choice of our scalings, the inverse of the parameters $\chi_\rho$, $D_\rho$ and $D_\eta$ can be interpreted as the P\'{e}clet numbers associated with $S$, $\rho$, and $\eta$, respectively. However, in this context $\chi_\rho$ has an additional meaning as it controls the time-scale at which the chemotactic signal converges to its chemical equilibrium. Specifically, small values of $\chi_\rho$ indicate a regime in which the profile of the chemotactic signal is quasi-steady, \emph{i.e.}, the time derivative in~\eqref{eq:burger_chemoattractant} can be neglected, yielding a chemical signal profile $S$ which results from the balance between transport and consumption of $S$. Eqs.~\eqref{eq:TW eq rho}-\eqref{eq:TW eq S} are supplemented by far-field conditions that determine the limiting behaviour of all variables at the back of the wave
    \begin{align}
        \rho\to 1, \quad \eta\to0,\quad S\to1, \quad z\to-\infty.    \end{align}
\end{subequations}
In writing~\eqref{eq:travelling_wave_population}, we have normalised the spatial and temporal dimensions respectively by $\ell=\tilde{D}_a/v$ and $\tau=D_a/v^2$ to balance the advective and diffusive transport of chemoattractant. Provided both the velocity $v$ and $\tilde{D}_a$ are known (as is the case for~\cite{Ucar2025}), this scaling choice reduces the total number of unknown parameters to six, where four are associated with unknown parameters, and two degrees of freedom are needed to fix the translational invariance of the travelling wave and the conserved mass of the sensor cells $\eta$. As also investigated in~\cite{Ucar2025}, solutions to the full model Eq.~\eqref{eq:time-dependent problem} supplemented with appropriate boundary conditions (Section A1, Supplementary Information) converge to travelling-waves described by Eq.~\eqref{eq:travelling_wave_population} as $t\to\infty$ (Figure S1).
\section{Weak chemoattractant sensitivities balance fast consumption during collective migration of dendritic and T cells}
\label{sec:fitting}
We calibrate Eq.~\eqref{eq:travelling_wave_population} to the experimental data from~\cite{Ucar2025} using Bayesian inference (Section B, Supplementary Information). The model recapitulates the spatial distribution of both populations measured experimentally (Figure~\ref{fig:fitting}a); additionally,  it reveals the spatial distribution of the chemotactic signal $S$ that could not be directly measured experimentally. Sampling from the inferred posterior probability density, we can use the model given in Eqs.~\eqref{eq:travelling_wave_population} to investigate the relation between the local strength of the chemotactic signal $S$ and the local density of dendritic and T cells (Figure~\ref{fig:fitting}b). Interestingly, we find that the posterior curves in the ($S$,$\rho$)-phase space collapse onto the curve $\rho\equiv S^2$, revealing an emergent square-power law scaling of the density $\rho$ with the signal $S$; this is despite the posterior having a broader uncertainty when looking at the full spatial distribution. Inspecting posterior samples for the value of the non-dimensional model parameter $(\chi_\rho,\kappa,D_\eta,D_\rho)$ (Figure S3 and Table S1, Supplementary information), we find that we confidently determine all non-dimensional parameters except $\chi_\rho$. Nonetheless, we can identify an upper bound and find strong statistical evidence that $\chi_\rho$ is significantly smaller than the other model parameters. The uncertainty in the estimate for $\chi_\rho$ propagates to the net consumption rate $\tilde{r}_\rho\rho^\dagger$ via~\eqref{eq:velocity} (\Cref{fig:additionalfittingC}, Supplementary Information), which must be sufficiently large to compensate for the small chemotactic sensitivity and to give a finite propagation speed. Physically, small values of $\chi_\rho$ indicate a regime in which the consumption dynamics is fast, reaching a quasi-steady approximation with shallow gradients that are limited by chemical diffusion. Consequently, the calibration should not be interpreted as providing a precise estimate for \(\chi_\rho\). Rather, the spatial data robustly identify the asymptotic regime \(\chi_\rho\ll D_\rho\), within which the front profile becomes insensitive to the precise value of \(\chi_\rho\).

We formally study the small-$\chi_\rho$ asymptotic limit (treating all other parameters as $\mathcal{O}(1)$) to shed light on the correlation between the signal and the population profiles observed in the calibrated model. The asymptotic analysis reveals that the signal $S$ takes a logistic shape
\begin{equation}
    S(z)\sim \frac{1}{\exp\left[\frac{z-z_0}{2 D_\rho}\right]+1},\label{eq:S explicit}
\end{equation}
where $z_0$ sets the arbitrary location of the wave front (without loss of generality, we set  $z_0=0$) and $2D_\rho$ is the front thickness which corresponds to the inverse of the consumer cell P\'eclet number, which indicates the spatial scale at which cell diffusion and directed migration balance. For the consumer populations, we recover the square power-law observed in the calibrated model (Figure~\ref{fig:fitting}) 
\begin{equation}
    \rho\sim S^2\sim\frac{1}{\left(\exp\left[\frac{z}{2 D_\rho}\right]+1\right)^2}. \label{eq:rho explicit}
\end{equation}
We note that the square-logistic type of profile in~\eqref{eq:rho explicit} is also a special solution for the standard Fisher-KPP model for a specific choice of the travelling wave speed~\cite{ablowitz_explicit_1979}; consistent with a diffusive interface front morphology. The sensor population also follows a well-known distribution relative to the chemical signal variable $S$:
\begin{equation}
    \eta\sim C_\eta\, S^{\alpha-1}(1-S)^{\beta-1},\label{eq:eta explicit}
\end{equation}
where the normalising factor $C_\eta=\Gamma(\alpha+\beta)/\Gamma(\alpha)\Gamma(\beta)$ is a positive constant, with $\Gamma$ indicating the Gamma function. In Eq.~\eqref{eq:eta explicit}, the shape parameters $(\alpha,\beta)$ are directly connected to the physical model parameters via
\begin{equation}
    \alpha=\frac{2D_\rho}{D_\eta}+1, \quad  \beta=\frac{2D_\rho(\kappa-1)}{D_\eta}+1.
\end{equation}
We therefore find that the sensor population follows a beta distribution with shape parameters $\alpha$ and $\beta$ determined by the relative diffusivities of the two populations $\nu=D_\eta/D_\rho$ and the relative chemotactic sensitivities $\kappa>1$ of the two cell populations. We can combine Eqs.~\eqref{eq:S explicit}-\eqref{eq:eta explicit} to study the spatial organisation of the two populations relative to each other. In particular, we focus on the localisation of sensor cells relative to the advancing consumer front. To this end, we use two key metrics to quantify the spatial arrangement; namely, the mean ($\mu$) and variance ($\sigma^2$) of the distribution $\eta(\rho)$, which is well defined given that the profile of $\rho$ is monotonic~\eqref{eq:rho explicit} (Figure~\ref{fig:fitting}). When looking at these quantities for the posterior model predictions, we find that $\mu\approx0.51$ and $\sigma^2\approx0.96\sigma^2_{\max}$, where $\sigma^2_{\max}=1/12$ is the maximum value that the variance can take (Section D1, Supplementary Information).~These estimates suggest that the system falls into a parameter regime where a larger fraction of the T cells are equally distributed at the front and back of the advancing group of consumer cells. Furthermore, the spread of the T-cell population is close to the maximal value predicted by the theory, as captured by the relatively high standard deviation. Interestingly, our theoretical analysis reveals how the system sits in a parameter regime in which the spatial organisation of the front is robust to perturbation in both key theoretical parameters $(\kappa,\nu)$; this is apparent when estimating the magnitude of the gradient in $\mu$ (Figure~\ref{fig:fitting}).~This suggests some level of robustness in the relative localisation of the two populations, while still allowing for flexibility in the spatial arrangement of the two populations, which is key to enabling us to identify model parameters. However, these results are limited to the small-$\chi_\rho$ regime. To understand robustness more generally, it is necessary to fully characterise solutions to~\eqref{eq:travelling_wave_population}. 

\section{A nonlinear theory of consumer/sensor migrating fronts}

The analysis in \parencite{Ucar2025} focuses on determining the migration speed, which is obtained through a linearisation far behind of the travelling wave front. As such, it does not provide information about the spatial organisation of the travelling front shaped by the consumer cells and the relative localisation of sensor and consumer cells within the migrating population.~While Ucar et al.~address this aspect by solving~\eqref{eq:time-dependent problem} numerically, an analytic estimate of travelling wave profiles would provide a more comprehensive understanding of how the nonlinear nature of chemoattractant signalling shapes the spatial structuring of sensor/consumer populations. Here, we address this gap by providing a full nonlinear theory of~\eqref{eq:travelling_wave_population} that describes the spatial organisation of two populations -- a consumer $\rho$ and a sensor population $\eta$, across a wider range of parameter regimes. Our theory fully captures how the spatial localisation of the two populations is shaped by the nonlinear nature of chemical interactions. This analysis allows us to extend the local results on robustness of the co-migration of T/DC cells obtained in Section~\ref{sec:fitting} to large variations in parameter values.

Briefly, we first identify the self-consistent dominant physical balances of~\eqref{eq:travelling_wave_population} corresponding to as many terms balancing at once (`distinguished asymptotic limits'). All system behaviours then either fall under one of the distinguished limits or as a `sublimit' of one of the distinguished limits. For each of the sublimits, we adopt asymptotic techniques to obtain analytic approximate solutions for all model variables~\cite{bender_advanced_1999}. While standard power series asymptotic expansions are sufficient to capture the structure of the solution in some sublimits, other sublimits require resolving the behaviour in different spatial regions via a matched asymptotic expansion approach, and the tracking of important but exponentially small terms. 

\begin{figure*}[t!]%[tbhp]
\centering
\begin{subfigure}{0.005\textwidth}
    \captionlistentry{}
    \label{fig:limit_interface_A}
\end{subfigure}
\begin{subfigure}{0.99\textwidth}
    \includegraphics[width=\linewidth]{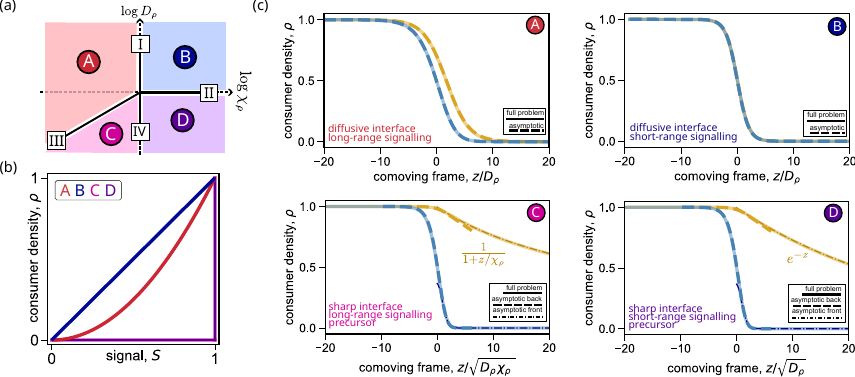}
    \captionlistentry{}
    \label{fig:limit_interface_B}
\end{subfigure}
\begin{subfigure}{0.005\textwidth}
    \captionlistentry{}
    \label{fig:limit_interface_C}
\end{subfigure}
\vspace{-7mm}
    \caption{\textbf{Characterisation of consumer front morphologies.} (a) The distinguished asymptotic limits of the TW problem~\eqref{eq:travelling_wave_population} for asymptotically large or small values of $\chi_\rho$ and $D_\rho$, denoted by dark solid lines and labelled using Roman numerals contained within squares. The distinguished limit $\rm{I}$ corresponds to $D_p\gg1$ with $\chi_\rho=\mathcal{O}(1)$, the distinguished limit $\rm{II}$ corresponds to $\chi_\rho\gg1$ and $D_p=\mathcal{O}(1)$, the distinguished limit $\rm{III}$ corresponds to $D_\rho/\chi_\rho= \mathcal{O}(1)$ with $D_\rho, \chi_\rho\ll1$ and the distinguished limit $\rm{IV}$ corresponds to $D_\rho\ll1$ and $\chi_\rho=\mathcal{O}(1)$. The sublimits of the distinguished asymptotic limits are denoted using letters contained within circles. (b) Leading-order prediction for the relationship between the signal $S$ and consumer population distribution in the different asymptotic regimes. Note that the pink and purple curves overlap. (c) Spatial distribution of the consumer cell $\rho$ (blue curve) and signal $S$ (golde curve) for different choices of model parameters, corresponding to the distinct regimes in Figure~\ref{fig:limit_interface_A}. We compare the numerical solution of~\eqref{eq:travelling_wave_population} (continuous curves) with the prediction from the asymptotic analysis (dashed lines). In the regimes C and D, two distinct asymptotic regions form: the outer front region and the inner region delineating the consumer interface. Computing the leading-order solution requires matching asymptotic corrections across the two regions (Section C, Supplementary Information). Values of the parameters used in panel C are listed in Table S2 (Supplementary Information).} 
\label{fig:phase diagram}
\end{figure*}

\subsection{Morphology of the consumer front}
\label{sec:front morphology}
We start by focusing on the coupled behaviour of the $(\rho,S)$ subsystem and construct the morphology diagram (Figure~\ref{fig:phase diagram}) characterising the distinct consumer front structures predicted by the model. We find that Eq.~\eqref{eq:travelling_wave_population} admits four distinguished asymptotic limits (curves identified with Roman numerals in Figure~\ref{fig:phase diagram}) depending on the relative asymptotic sizes of $D_\rho$ and $\chi_\rho$. Crossing these curves alters the dominant balance between the different physical mechanisms guiding migration, leading to a substantial change in the structure of the solution. As indicated in Figure~\ref{fig:phase diagram}, these curves separate four different regions of parameter space (indicated with capital letters) corresponding to the asymptotic sublimits.  Notably, while the travelling wave speed~\eqref{eq:travelling_wave_population} is independent of $D_\rho$, this parameter is key to determining its spatial structure. The sublimits correspond to significant differences in the morphology of the consumer front, \emph{e.g.}, the sharpness of the consumer interface and the signal propagation ahead of the front. 

We use asymptotic analysis to formally approximate the leading-order (dominant) behaviour of $\rho$ and $S$ in each of these sublimits. While the obtained solution in regimes A and B is valid across space, there is no single physical balance that defines the structure of the solution in regimes C and D. Rather, the structure of the solutions in regimes C and D is a composite of different asymptotic regions, which are only valid in specific regions of space that we identify as part of the solution. Different asymptotic regions emerge as the random cell motion becomes negligible, \emph{i.e.}, conditions where the cell P\'eclet number is smaller than the chemoattractant P\'eclet number. As a result, the balance of the chemoattractant dynamics across the interface is different from that in the outer region in front of the wave. Resolving these regimes is more involved since the balance between the different mechanisms changes across the consumer front, with the formation of a boundary layer across the interface formed by consumer cells. Hence, the method of matched asymptotics is required to correctly connect the solution at the front and rear of the travelling wave (see Section C, Supplementary Information). The boundary layer, whose size scales with $\tilde{\varepsilon}=\sqrt{\chi_\rho/(\chi_\rho+1)D_\rho}$, is the region of space where the random and directed motion of consumer cells balance. Taking the limit of $\tilde{\varepsilon}\to 0$ therefore corresponds to $\rho$ formally converging towards a Heaviside (step) function. In these regimes, the distribution of the signal extends ahead of the wave, highlighting how sharp fronts can maintain chemotactic influence on other cells over long distances and in the absence of direct cell–cell contact. Below, we briefly discuss the important physical and qualitative differences in each sublimit.

\subsubsection*{Regime A: $\chi_\rho \ll 1$, $D_\rho \gg \chi_\rho$} In this regime, chemical diffusion is negligible and the explicit solution~\eqref{eq:S explicit}-\eqref{eq:rho explicit} is obtained by balancing consumption and advective transport of the chemical signal $S$, leading to $\rho \sim S^2$. Notably, provided that the characteristic length scales of chemical signalling remains smaller than that of the interface, \emph{i.e.}, $D_\rho \gg \chi_\rho$, this regime remains valid across a broad range of parameter values, including regimes of both small and large cell diffusivities. As a result, regime A occupies a comparatively large region of parameter space, suggesting that the consumer front morphology is particularly robust to perturbations in model parameters, especially variations in $D_\rho$. 
\subsubsection*{Regime B: $\chi_\rho, D_\rho \gg 1$} Transitioning between regimes A and B maintains the global structure of the solution to~\eqref{eq:travelling_wave_population}, which is obtained by balancing the chemical consumption in the moving reference frame. This yields a strong coupling between $\rho$ and $S$; specifically, the density of consumer cells $\rho$ directly scales with the signal $S$ and both follow a sigmoidal curve
\begin{equation}
    \rho\sim S\sim \frac{1}{e^{z/D_\rho}+1}.\label{limitB:S&rho}
\end{equation}
Here, the characteristic interfacial thickness is equal to the inverse P\'eclet number $D_\rho$, which corresponds to the length scale at which directed and random motion of cells balance. In this regime, the propagation of the signal is short-range and therefore only cells within the consumer front remain chemotactically coupled to the signal. Unlike for regime A, this physical balance strictly requires the diffusion of the cells to be significantly larger than the chemoattractant diffusivity ($D_\rho \gg 1$). This asymptotic regime corresponds to a large wave speed approximation (as originally developed by Canosa~\cite{canosa}) often used in the context of travelling wave analysis.

\subsubsection*{Regime C: $D_\rho \ll \chi_\rho \ll 1$} In this regime, both the chemoattractant and cell P\'{e}clet numbers are large, and the interfacial thickness is set by $\tilde{\varepsilon}=\sqrt{\chi_\rho D_\rho}$. Unlike in regime A, chemoattractant consumption cannot balance the signal transport since the consumer cell density decays much faster, yielding exponentially small values of $\rho$ only a short distance ahead of the front (dark blue dashed line Figure~\ref{fig:limit_interface_C})
\[\rho\sim \frac{1}{e}\exp\left[-\frac{z^2}{D_\rho\chi_\rho}\right],\quad z>0.\]
Since $\rho$ is exponentially small ahead of the wave, the corresponding chemoattractant distribution is shaped by chemoattractant diffusion and advection (which require sharp gradients $\partial_zS\sim1/\chi_\rho$)
\[S\sim \frac{\chi_\rho}{\chi_\rho+z}, \quad z>0.\]
The algebraic decay of the signal ahead of the wave allows distant cells to remain chemotactically coupled to the consumer front resulting in the possibility of co-migration without the need for physical contact.
In contrast, local chemoattractant consumption becomes important across the interface and at the back of the wave, where the size of the consumer population $\rho = \mathcal{O}(1)$. At the back of the wave, steady reactions and signal transport balance as $\rho,S\sim 1$. Across the interface, instead, the chemoattractant P\'eclet number is small, indicating that diffusion of the chemoattractant is dominant, resulting in shallow gradients compared to the interfacial lengthscale; specifically, $\partial_uS\sim \sqrt{\chi_\rho/D_\rho}$ where $u=x/\tilde{\varepsilon}$ is the inner variable.

\subsubsection*{Regime D: $\chi_{\rho} \gg 1$, $D_{\rho} \ll 1$} In this regime, the behaviour of the solution across the interface is largely the same as in regime C, but with a slightly modified interfacial width ($\tilde{\varepsilon}=\sqrt{D_\rho}$). The major difference in the solution structure is the spatial distribution of the chemotactic signal in front of the wave. Since $\chi_\rho\gg1$, the profile of $S$ is set by equilibrating diffusion in the reference frame of the consumer front, which yields an exponential decay
\begin{equation}
S\sim e^{-z},\quad z>0. \label{eqD: signal front}
\end{equation}
The rate of decay of the signal is slower than that of the consumer cell density (dark blue dashed line Figure~\ref{fig:limit_interface_C})
\begin{equation}
\rho\sim \frac{1}{e}\exp\left[-\frac{z-1+e^{-z}}{D_\rho}\right]
\end{equation}
justifying why the leading-order behaviour of $S$ is independent of cell consumption in this asymptotic region. 
The exponential (rather than algebraic) decay in the chemotactic signal suggests that the range of signal influence in regime D decays much faster than in regime C. Nonetheless, in both regimes cells away from the consumer interface can be coupled to it via chemical signalling.

\subsection{Colocalisation of consumer-sensor populations}

We now use the analytical nonlinear theory we have developed for the consumer front to quantify the colocalisation (or overlap) of the consumer and sensor populations, and investigate how this changes with increased heterogeneity in cellular behaviour. Biologically, colocalisation can be critical for mediating cell functions such as immune response, where physical proximity between cell types is often required for effective function. Our analytic results will allow us to generalise the numerical investigation in~\cite{Ucar2025} to a wider range of parameters and gain insights into the mechanisms driving mixing of consumer-sensor cells during migration.

\begin{figure*}
    \centering
    \includegraphics[width=\linewidth]{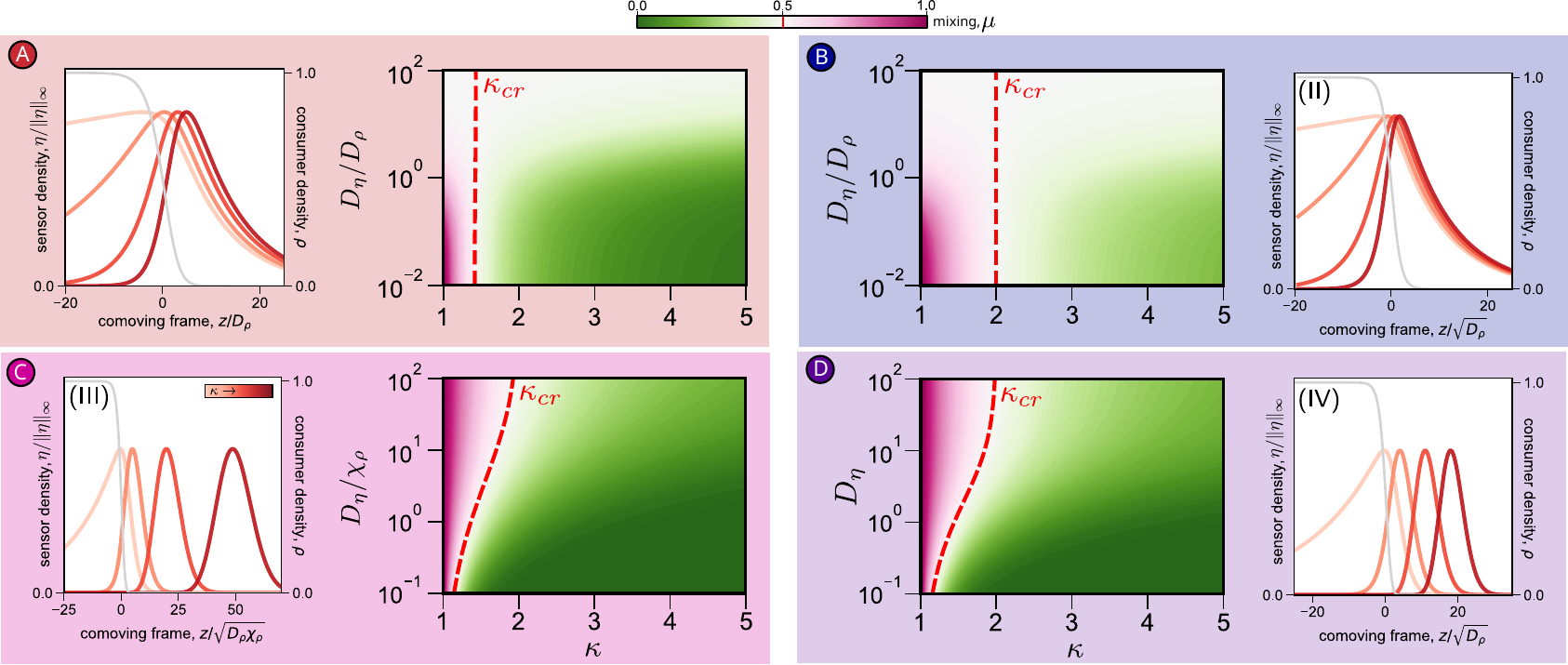}
    \caption{Diagrams illustrating how colocalization of consumer-sensor populations depends on the level of heterogeneity in the migratory properties of the two cell types for the different consumer front morphologies A-D (defined as in Fig.~\ref{fig:phase diagram}). The colormap indicates the magnitude of the mixing metric~\eqref{eq:general mixing}. The red dashed line indicates maximal mixing conditions (corresponding to $\mu=0.5$). (I)-(IV) Spatial distribution of the sensor cells (red curves) for increasing values of $\kappa$ -- sensor-to-consumer chemotactic sensitivity -- for the different regimes. The consumer profile is the same as in Figure~\ref{fig:phase diagram}c, $D_\eta/D_\rho=10$ and $\kappa\in\left\{1.05,1.5,3.0,6.0\right\}$.}
    \label{fig:mixing}
\end{figure*}

We extend the analysis from Section~\ref{sec:fitting} to quantify how the spatial organisation of sensor cells across the consumer front depends on its migratory properties -- relative to those of the consumer. Specifically, we are interested in investigating the mechanisms that allow for robust co-migration of the two cell populations. Numerical simulations reveal that the consumer front morphology fundamentally changes how sensor-consumer heterogeneity affects the location of sensor cells within the migrating collective (Figures~\ref{fig:mixing}$\sf{I}$-$\sf{IV}$). For example, up to a 6-fold increase in the sensor cell responsiveness to the chemoattractant, $\kappa$ easily disrupts the coherent formation of migrating cells, yielding complete separation in the sharp interface regime (Figures~\ref{fig:mixing}$\sf{III}$-$\sf{IV}$) while still allowing mixing of consumer and sensor cells in the diffusive regime (Figures~\ref{fig:mixing}$\sf{I}$-$\sf{II}$). In this section, we aim to use the theory derived in Section~\ref{sec:front morphology} to understand how heterogeneity in consumer-sensor systems affects their ability to coherently migrate.

Briefly, we obtain the leading-order behaviour of $\eta$ by appropriately integrating~\eqref{eq:TW eq eta} using the relevant expression for the signal function $S$ along the consumer front (Figures~\ref{fig:mixing}$\sf{I}$-$\sf{IV}$). As in Section~\ref{sec:fitting}, we then quantify the spatial overlap between the two populations via the mixing metric
\begin{equation}
    \mu=\int_0^1 \rho \,\mathrm{d} F_\eta(\rho),\label{eq:general mixing}
\end{equation}
where the probability measure $F_\eta(\rho)$ represents the fraction of sensor cells located at a spatial location where the consumer density is $\leq\rho$, as determined by the leading-order approximation of the cell densities. The metric $\mu$ interpolates between complete separation of the two populations ($\mu=0$) and complete overlap ($\mu=1$). The threshold value $\mu=0.5$ indicates the transition between these two regimes, whereby the sensor population switches from being positioned primarily ahead of the consumer cells to accumulating at the back of the consumer front.

We find that $F_\eta$ takes the form of well-known probability distributions and, therefore, we can estimate~\eqref{eq:general mixing} as the moments of known probability distributions (Section D, Supplementary Information). As shown in Figure~\ref{fig:mixing}, the dependence of the colocalisation on the relative properties of the sensor and consumer population significantly changes depending on the shape of the migrating front. In regimes A and B, owing to the strong spatial coupling between the consumer cell distribution and the chemoattractant signalling, the mixing only depends on the relative migratory properties of the sensor and consumer population. In contrast, in regimes C and D, mixing depends differently on the sensor cell diffusivity. The presence of long-range distribution of the chemical signal in the sharp interface case -- namely, regimes C and D -- facilitates the demixing of the two populations, allowing sensor cells to migrate even if spatially segregated from the consumer cells, especially for low values of $D_\eta$. 

\subsection*{Diffusive consumer front} When the consumer front is diffusive (regimes A and B), the signal distribution $S$ and the consumer density profile $\rho$ are strongly coupled. As a result, the spatial arrangement of sensor cells is solely determined by its properties relative to the consumer cells. 

In regime B, we find that $\mu$ is the mean of a \emph{Beta} distribution where the corresponding shape parameters depend on the relative diffusivities $D_\eta/D_\rho$ and the chemotactic sensitivity $\kappa$ of the two populations (Section D2, Supplementary Information). While the critical value $\kappa_{cr}$ for which $\mu=0.5$ is independent of random motion ($\kappa_{cr}=2.0$), this still affects the mixing of the two populations; specifically, large values of $D_\eta/D_\rho$ also increase mixing as $\mu\to0.5$; the limit of small $D_\eta/D_\rho$ instead yields $\mu\sim\kappa^{-1}$ and hence corresponds to the regime in which mixing is most sensitive to $\kappa$. 

When comparing with the estimates of $\mu$ for regime A (Section~\ref{sec:fitting}), we find that $\mu_A<\mu_B$ given the same value of $\kappa$ and relative diffusivities $D_\eta/D_\rho$. This is consistent with the results from Section~\ref{sec:front morphology}; the short-range nature of the chemoattractant signal $S$ in regime B results in a larger level of mixing compared to regime A, for which the chemoattractant signal extends over a longer range (Figure~\ref{fig:limit_interface_C}), and therefore chemotaxis of the sensor population is less dependent on physical proximity to consumer cells. However, shorter-range signalling impairs sensor cells' ability to lead the migrating collective. Specifically, in regime B, sensor cells require approximately a two-fold increase in chemotactic sensitivity to localise effectively ahead of the wave. In contrast, only an approximately 1.4-fold increase is required in regime A. Hence, for weakly-heterogeneous populations, such as for the dendritic and T-cells mixture analysed in Section~\ref{sec:fitting}, longer-range signalling may be selected to balance localisation of sensor cells at the leading edge of the migrating collective with maintenance of a coherent mixed migrating population. This further supports our findings that the dendritic-T cell system operates within a region of parameter space that enhances robustness of the spatial structure of the migrating front.

\subsection*{Sharp consumer front} In the sharp-interface limit (regimes C and D), under the additional assumption that $D_\eta\gg D_\rho$ (Sections D3-D4, Supplementary Information), the leading-order behaviour of $\rho$ is given by a Heaviside (step) function and $F_\eta$ is equivalent to the cumulative distribution of a Bernoulli random variable, and therefore the corresponding $\mu$ is equivalent to the Bernoulli success probability. Essentially, in the sharp-interface limit localisation is effectively reduced to a binary partition between cells remaining behind the interface and cells migrating ahead of it. 

Generally, the relationship between the estimated mixing metric $\mu$ and model parameters differs in the two regimes (Sections D3-D4, Supplementary Information), yet they display the same limiting behaviour in the large-$D_\eta$ limit with $\mu\to\kappa^{-1}$. This is equivalent to the behaviour of $\mu$ in regime B in the small-$D_\eta/D_\rho$ limit. Hence, in contrast to regimes A-B, even for large values of sensor motility $D_\eta\gg1$, the level of mixing between the two populations remains sensitive to the parameter $\kappa$ and the transition to a sensor-lead migrating front ($\mu\approx0.5$) is less smooth. For example, small variations in $\kappa$ can result in significant rearrangements of sensor cells and their detachment from the consumer front. 

\section*{Discussion}
Understanding how heterogeneous cell populations coordinate migration through self-generated chemotaxis is a central problem in collective cell dynamics. Here, we extended our theoretical understanding of this process within the non-reciprocal sensor-consumer systems~\eqref{eq:time-dependent problem} developed by~\cite{Ucar2025}. Systematic calibration of model~\eqref{eq:travelling_wave_population} to experimental data from migrating mixtures of dendritic and T cells reveals that spatial information alone is sufficient to recover all model parameters except one and the asymptotic balance shaping their collective migration, even without direct measurements of the chemoattractant profile. This is possible because information about the chemotactic signal is encoded in the nonlinear relationship between the spatial arrangement of dendritic and T cells and the morphology of the advancing dendritic (\emph{i.e.}, consumer) front.

We therefore derive a nonlinear theory of sensor-consumer migrating fronts driven by chemotaxis through systematic asymptotic analysis of~\eqref{eq:travelling_wave_population}. This allows us to disentangle how model parameters regulate the front morphology (Figure~\ref{fig:phase diagram}) and robust collective migration of a mixture of chemotacting heterogeneous cell types. Our findings broadly agree with previous observations from~\cite{Ucar2025} based on numerical explorations and linear analysis of~\eqref{eq:travelling_wave_population}. Importantly, our analysis shows that robust collective migration can persist even in the limit in which cellular diffusion is negligible relative to chemoattractant diffusion ($D_\rho,D_\eta\ll1$). We find that the dendritic cells fall in the region of parameter space that leads to diffusive front interfaces and intermediate chemotactic signalling (regime A in Figure~\ref{fig:phase diagram}).~This regime spans biologically plausible scenarios ranging from migration along diffusible cues ($D_\rho\ll1$) to haptotactic guidance by substrate-bound signals -- which have a small diffusion coefficient ($D_\rho\gg1$)~\cite{patel_chemokines_2001}, provided chemoattractant consumption and transport is equilibrated during migration ($\chi_\rho\ll D_\rho$). The finite, yet extended, range of chemotactic signalling facilitates localisation of sensor cells at the leading edge while preserving formation of a coherent mixed front. This contrasts with sharp-interface regimes, where long-range and diffusive signalling ahead of the consumer front promotes spatial separation between the sensor and consumer populations and weakens coherent collective organisation. 

More broadly, our framework provides a general approach for understanding robust collective migration in heterogeneous populations driven by self-generated chemotaxis. Beyond the dendritic/T-cell system considered here, the theory can be applied to a wide range of biological and synthetic systems exhibiting asymmetric chemical interactions. 
To isolate the role of asymmetric chemical interactions and motility in shaping front organisation, we intentionally neglected additional mechanisms known to influence collective migration, including consumption of the chemoattractant by sensor cells~\cite{marjorie2026}, cell density-dependent consumption rates~\cite{phan_direct_2024}, mechanical interactions~\cite{ford_pattern_2025,celora2026chemotaxiscellaggregatesmorphology,jewell2026cellcelladhesionsustainextended,lawson-keister_collective_2022,rossetti_optogenetic_2024,panigrahi_intermittent_2025}, short memory effects in chemoattractant sampling~\cite{dolak_kinetic_2005,saragosti_mathematical_2010} and geometry of the fronts~\cite{alert_cellular_2022,bhattacharjee_chemotactic_2022}. Overall, our work demonstrates how asymptotic analysis combined with quantitative experimental calibration can uncover the physical principles linking signalling structure, front morphology, and robust collective migration in heterogeneous cell populations.

\section*{Acknowledgments}
The authors thank Mehmet Can U\c{c}ar for helpful discussions. For the purpose of Open Access, the authors have applied a CC BY public copyright licence to any Author Accepted Manuscript (AAM) version arising from this submission. GLC and CF thank financial support from the Mathematical Institute, University of Oxford, via the Hooke Research Fellowship. MW acknowledges support by funding from the UKRI-EPSRC (grant number EP/Y034791/1).

\section*{Data and Code Availability}

The code used to solve the model is available upon request and will be made publicly available upon publication. All model and simulation parameters used to generate the main text figures can be found in~\Cref{tab:estimates parameters}, Supplementary Information.

\printbibliography
\onecolumn
\appendix
\addtocontents{toc}{\protect\setcounter{tocdepth}{2}} % show subsections

\setcounter{figure}{0}    

\setcounter{equation}{0}    
\renewcommand*{\thesection}{\Alph{section}}
\renewcommand*{\thefigure}{S\arabic{figure}}
\renewcommand*{\thetable}{S\arabic{table}}
\renewcommand*{\theequation}{S\arabic{equation}}
\hspace{-.5cm}\large{\textbf{Supplementary Information}}
\vspace{1cm}
\normalsize
\tableofcontents
\pagenumbering{gobble}

\newpage
\pagenumbering{arabic}
\setcounter{page}{1}

\newpage
\section{Details of the SC model}
Starting from Eq.~(3) in the main text, we non-dimensionalise the modelling equations by introducing the following scalings
\begin{equation}
    \tilde{\rho}=\rho \rho^\dagger, \quad \tilde{S}=\frac{v}{\tilde{\chi}_\rho} S, \quad \tilde{x}=\frac{D_a}{v} x, \quad \tilde{t}=\frac{D_a}{v^2} t,\label{eq:scalings}
\end{equation}
where the tilde symbol $\tilde{\varphi}$ indicates the dimensional form of the variable $\varphi$ and $v$ is as defined in Eq.~(2) of the main text. Using the scalings~\eqref{eq:scalings} into Eq.~(3) in the main text, we find:
\begin{subequations}\label{sys:nondimensional}
    \begin{align}
    \partial_t\rho&=\partial_x(D_\rho\partial_x\rho-\rho S),\label{eq:rho transport pde}\\
    \partial_t\eta&=\partial_x(D_\eta\partial_x\eta-\kappa\eta S),\label{eq:eta transport pde}\\
    \partial_t S&=\partial_x\left(\partial_xS +\frac{S^2}{\chi_\rho}\right)-\frac{1+\chi_\rho}{\chi_\rho}\partial_x\rho,\label{eq:S transport pde}
\end{align}
\end{subequations}
where we have introduced the three non-dimensional parameters:
\begin{equation}
    \chi_\rho=\frac{\tilde{\chi}_\rho}{D_a},\quad D_\rho=\frac{\tilde{D}_\rho}{D_a}, \quad D_\eta=\frac{\tilde{D}_\eta}{D_a}, \quad \kappa=\frac{\tilde{\chi}_\eta}{\tilde{\chi}_\rho}.
\end{equation}
Looking at Eq.~\eqref{eq:S transport pde}, we see that the non-dimensional parameter $\chi_\rho$ controls the balance between advection and diffusion of the chemotaxis signal $S$, as well as the time-scale at which $S$ relaxes towards its equilibrium. Specifically, the small-$\chi_\rho$ limit corresponds to a quasi-steady approximation, whereby chemoattractant advection is balanced by consumption. In contrast, the large-$\chi_\rho$ limit implies a small P\'{e}clet number for $S$, and its profile is shaped by a balance between diffusion and non-equilibrium reaction dynamics. 

\subsection{Simulation of the dendritic/T-cell system}
\begin{figure}
    \centering
    \includegraphics[width=\linewidth]{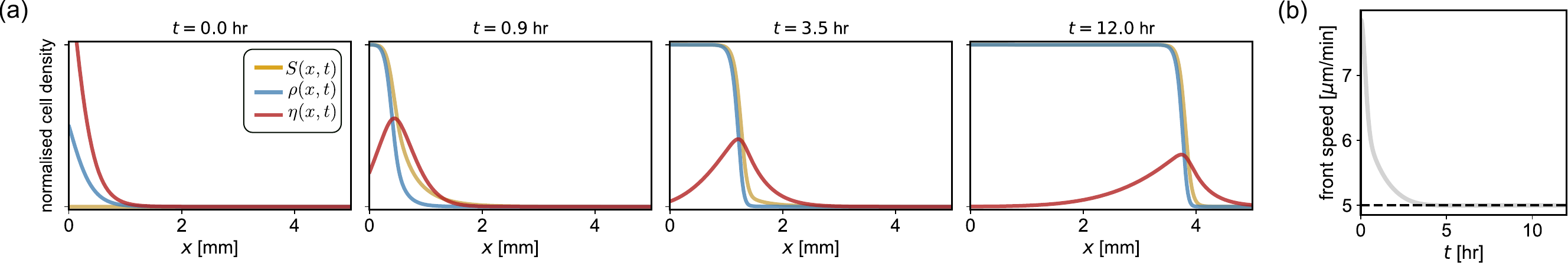}
    \caption{Simulation of the system~\eqref{sys:nondimensional}-\eqref{eq:initial conditions} for calibrated values of the model parameters in~\Cref{tab:estimates parameters}. To facilitate the numerical simulations, we take $\chi_\rho$ to be larger than the estimated value $\chi_\rho=5\times10^{-4}$, yet sufficiently small not to affect the model solution (see discussion in~\Cref{sec:calibration results}). Dimensional forms of the dependent and independent model variables are obtained by applying the scalings in~\Cref{dimensional scalings}. (a) Snapshot of the spatial distribution of the consumer $\rho$ and sensor $\eta$ populations, and chemotactic signal profile $S$. (b) Time evolution of the migrating front, defined as the spatial location $x_F$ at which $\rho(x_F(t),t)=0.5$. The numerical solution is compared to the expected asymptotic value of the migration speed $v=5\mu \text{m/min}$ and is equivalent to the one obtained experimentally in~\cite{Ucar2025}. }
    \label{fig:dynamic simulation}
\end{figure}

Following~\cite{Ucar2025}, we compare the SC model to the microfluidic system solving~\eqref{sys:nondimensional} on a large finite domain $x\in[0,L]$. At the outlet ($x=L$), we take no-flux conditions for the cell densities and a Neumann condition for the signal
\begin{equation}
    D_\rho\partial_x \rho(L,t)-\rho(L,t)S(L,t)=D_\eta\partial_x \eta(L,t)-\kappa\eta(L,t)S(L,t)= \partial_xS(L,t)=0.\label{eq: BC inlet}
\end{equation}
We take $L\gg1$ so that the choice of the behaviour at the outlet does not influence the migration dynamics. At the inlet, we impose no-flux conditions for the sensor populations, Neumann conditions for the chemotactic signal, and a non-zero influx for the consumer populations:
\begin{align}
       D_\rho\partial_x \rho(0,t)-\rho(0,t)S(0,t)=-F_\rho, \quad D_\eta\partial_x\eta(0,t)-\kappa\eta(0,t)S(0,t)=0,\quad \partial_xS(0,t)=0.\label{eq: BC outlet}
\end{align}
Here, $F_\rho$ is a non-dimensional cell influx normalised by the characteristic influx $\rho^\dagger v$. In contrast to~\cite{Ucar2025}, we do not consider no-flux boundary conditions for the chemoattractant, but rather assume that the signal is flat at the boundaries of the domain. This prevents the formation of boundary layers at the inlet, which are not observed in the experiments. This choice better captures the fact that consumer cells are also located at the left of the inlet and can therefore generate chemical gradients themselves. We initialise the system by setting an initial homogeneous chemoattractant concentration (\emph{i.e.}, no signal), and a truncated sigmoidal profile for the cell populations; specifically,
\begin{equation}
    \rho(x,0)=\eta(x,0)=\frac{2}{1 + \exp(\lambda_0x)}, \quad S(x,0)\equiv0,\label{eq:initial conditions}
\end{equation}%
with $\lambda_0=5$.~\Cref{fig:dynamic simulation} illustrates a characteristic simulation of~\eqref{sys:nondimensional}-\eqref{eq:initial conditions} given the estimated values of model parameters (see~\Cref{tab:estimates parameters} and~\Cref{sec:calibration}). Details on the numerical scheme used can be found in~\Cref{sec: timedependentsimulations}. We find that after a short transient, the solution quickly relaxes to a travelling wave profile. The timescale needed to converge to the corresponding travelling wave necessarily depends on the choice of the initial condition. For~\eqref{eq:initial conditions}, we find that within the timescale of the experiments, $\approx 6$ hours, the solution has already converged to its corresponding travelling wave profile and the front advances with the expected travelling wave speed $v$.
\subsubsection{Numerical Scheme}
\label{sec: timedependentsimulations}
We solve Eqs.~\eqref{sys:nondimensional}-\eqref{eq:initial conditions} using the method of lines. In particular, we discretise the spatial interval $x\in[0,L]$ into $N_x$ subintervals $C_j$ of size $\Delta_x$: $C_j=[x_j,x_{j+1}]$ $x_j=j\Delta x$. We adopt a mixed scheme, where we solve for $S$ at the boundary of each mesh interval $S(t,x_j)\approx S_j$, while the cell densities are estimated at the centre of the domain via a finite-volume approximation:
\begin{equation}
    S(t,x_j)\approx S_j, \quad j=0,\ldots,N, \quad \frac{\int_{C_j}\rho(t,y)dy}{\Delta x}\approx \rho_{j+1/2}, \quad j=0,\ldots,N-1.
\end{equation}%
To discretise the transport equation~\eqref{eq:S transport pde}, we adopt a central difference for the diffusion operator and upwind scheme (given a negative speed) for the nonlinear advection term:
\begin{equation}
    \frac{\d S_i}{\d t}=\frac{S_{i+1}+S_{i-1}-2S_{i}}{\Delta x^2}+\frac{1}{\chi_\rho}\frac{S^2_{i+1}-S^2_i}{\Delta x}-\frac{1+\chi_\rho}{\chi_\rho}\frac{\rho_{i+1/2}-\rho_{i-1/2}}{\Delta x},\quad i=1,\ldots,N-1, \label{eq:discretised S}
\end{equation}
while the boundary terms are set by strongly imposing the Neumann conditions: $S_0=S_1$ and $S_N=S_{N-1}$. To obtain the evolution of $\rho_j$ and $\eta_j$, we integrate~\eqref{eq:rho transport pde} and \eqref{eq:eta transport pde}, respectively, over the interval $C_i$:
\begin{equation}
    \frac{\d\varphi_j}{\d t}=\left[\varphi(x,t) v_\varphi(x,t)\right]^{x_{j+1}}_{x_j}, \quad v_\varphi(x,t)=D_\varphi\partial_x\log\varphi-\frac{\chi_{\varphi}}{\chi_\rho}S,
\end{equation}
for $j=1,\ldots,N-1$ and $\varphi=\rho,\eta$.
Next, we need to approximate the fluxes $\mathcal{F}_\varphi=\varphi v_\varphi$ at the cell edges. For the diffusive contribution $\partial_x \log\varphi$ to the velocity, we adopt a central finite difference
\begin{equation}
    \partial_x\log(\varphi)\approx \frac{\log(\varphi_{j+1/2})-\log(\varphi_{j-1/2})}{\Delta x},
\end{equation}
while the chemotactic contribution to the velocity is readily approximated by $S_j$, so that
\begin{equation}
    v^{j}_\varphi=D_\varphi\frac{\log(\varphi_{j+1/2})-\log(\varphi_{j-1/2})}{\Delta x}-\frac{\chi_{\varphi}}{\chi_\rho}S_j, \quad j=1,\ldots,N-1.
\end{equation}
We then construct the cell flux at the inner cell edges using an upwind approximation
\begin{equation}
    \mathcal{F}_\varphi^j=\varphi_{j+1/2}(v_\varphi^j)^-+\varphi_{j-1/2}(v_\varphi^j)^+, j=1,\ldots,N-1, \quad \varphi=\rho,\eta,
\end{equation}
where $(v)^+=\max(0,v)$ and $(v)^-=\min(0,v)$. The flux at the boundary edges is instead prescribed via the boundary conditions~\eqref{eq: BC inlet}-\eqref{eq: BC outlet}; specifically, $\mathcal{F}_\rho^0=F_\rho$ and $\mathcal{F}_\eta^0=\mathcal{F}_\rho^N=\mathcal{F}_\eta^N=0$. Hence, we obtain the system of ODEs:
\begin{equation}
    \frac{d\varphi_{j}}{dt}=\frac{\mathcal{F}_{j+1/2}-\mathcal{F}_\varphi^{j-1/2}}{\Delta x}, \quad j=\frac{1}{2},\ldots,N-\frac{1}{2}, \quad \varphi=\rho,\eta.\label{eq: discretised cell density}
\end{equation}
We integrate the coupled systems of ODEs~\eqref{eq:discretised S} and~\eqref{eq: discretised cell density} numerically using the time-stepping Dormand–Prince 5th order Runge–Kutta method — implemented in \texttt{scipy.integrate} — with a limit on the maximum timestep to ensure the system properly converges to its corresponding travelling wave with the appropriate travelling wave speed. 

\subsection{Reduction to a travelling wave problem}
We seek travelling wave solutions to~\eqref{sys:nondimensional} by looking for solutions of the form:
\begin{equation}
    (\rho,\eta,a)=(\rho(z),\eta(z),a(z)), \quad z=x-t, \label{eq: TW ansatz}
\end{equation}
where the travelling velocity is set to $1$ in line with the scalings~\eqref{eq:scalings}. Substituting~\eqref{eq: TW ansatz} into~\eqref{sys:nondimensional}, we obtain the system of coupled boundary value problems
\begin{subequations}\label{sys: TW 1}
\begin{align}
    (D_\rho\rho'+\rho-S\rho)'=0,\\
    (D_\eta\eta'+\eta-\kappa S\eta)'=0,\\
    \left(S'+\frac{S^2}{\chi_\rho}+S-\frac{1+\chi_\rho}{\chi_\rho}\rho\right)'=0,
\end{align}
with far-field conditions connecting the two steady states of~\eqref{sys: TW 1}
\begin{align}
    \lim_{z\to-\infty}(\rho,\eta,S)=(1,0,1),\quad \lim_{z\to\infty} (\rho,\eta,S)=(0,0,0).\label{eq: far-field}
\end{align}
\end{subequations}
Owing to the conservative structure of~\eqref{sys: TW 1}, we can integrate each equation and, by imposing any of the two far-field conditions, we obtain the following system of coupled first-order ODEs
\begin{subequations}
\begin{align}
    D_\rho\rho'&= \rho\left(S-1\right),\\
    D_\eta\eta'&=\eta(\kappa S-1),\\
    \chi_\rho S'&=\rho(\chi_\rho+1)-S\left(S+\chi_\rho\right),
\end{align}\label{sys:fitting}%
with far-field conditions~\eqref{eq: far-field}. Note that the two far-field conditions are consistent with one another owing to the integration of~\eqref{sys: TW 1}; hence, any of the two can be applied.
\end{subequations}
While the dynamics of $\eta$ decouple from those of $\rho$ and $S$, the profile of $\eta$ helps with identifying model parameters as it contains information on the experimentally hidden signal distribution. Hence, in the fitting, we estimate all parameters simultaneously. 

\section{Calibration of the SC model to the experimental data for the dendritic/T-cell system}
\label{sec:calibration}
\begin{figure}[htb]
    \centering
    \begin{subfigure}{0.0\textwidth}      \captionlistentry{}
    \label{fig:experimental profile lab frame}\end{subfigure}\includegraphics[width=\linewidth]{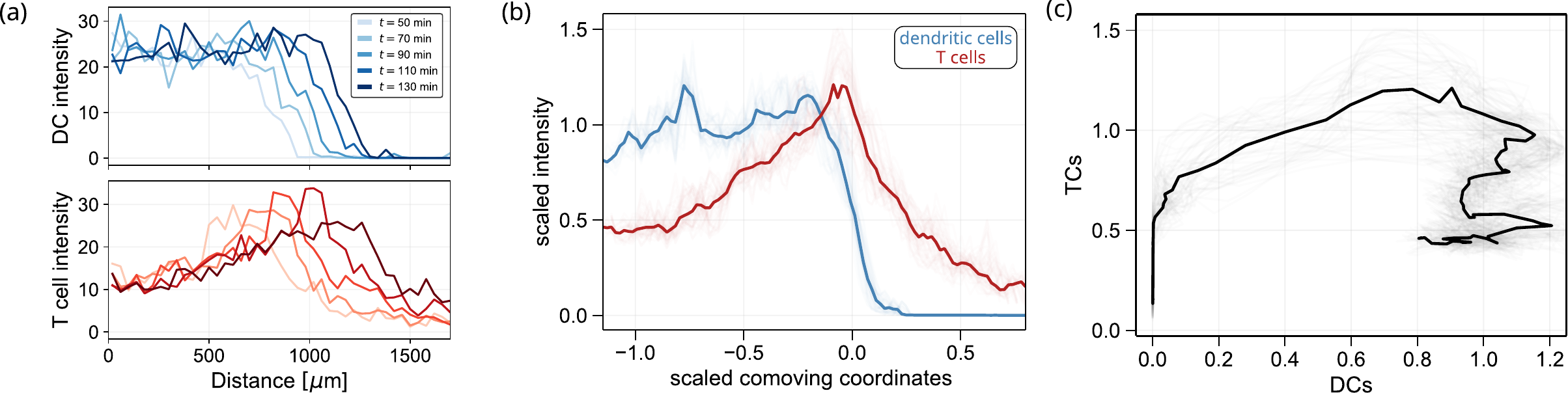}
    \caption{(a) Spatio-temporal evolution of the dendritic (DC) and T (TC) cells intensity from experiments~\cite{Ucar2025}. (b) Normalised spatial distribution of cell populations in the co-migration reference frame.~Dark lines indicate the mean distribution. (c) Projection of the experimentally estimated travelling wave profile in the dendritic-T cells plane. Panels (a)-(b) are adapted from Figure 3 in~\cite{Ucar2025} (Obtained from~\cite{Ucar2025}, licensed under CC BY 4.0).}
    \label{fig:experimental data}
\end{figure}

We fit~\eqref{sys:fitting} to microscopy images from the microfluidic channel experiments in~\cite{Ucar2025}. The authors extrapolate the spatiotemporal cell density profiles, labelling cells with different fluorescent markers and observing them as they migrate across the microfluidic channels (Figure~\ref{fig:experimental profile lab frame}). The chemoattractant gradient, and, therefore, the chemical signalling variable, is not measured. The experimentally estimated travelling speed is $v=5\,\mu\text{m/min}$. The spatial rescaling is then based on the known diffusion coefficient of the chemoattractant $D_a=86\,\mu\text{m}^2/\text{s}$~\cite{Ucar2025}. Hence, based on~\eqref{eq:scalings}, we have the following characteristic temporal and spatial scales
\begin{equation}
T=\frac{D_a}{v^2}\approx 206.4\text{ min},\quad X=\frac{D_a}{v}\approx1\text{ mm}.\label{dimensional scalings}
\end{equation}

To obtain the estimates of the travelling-wave profile, we overlay the cell density profiles from the final 10 experimental measurements in the comoving frame. Under the assumption that the system is relaxed to its travelling wave profile, we used the overlaid profiles to obtain an estimate of the cell spatial distribution and of the uncertainty in the estimates. Following this procedure, we obtain the normalised profiles illustrated in~\Cref{fig:experimental data}. The consumer cell density is normalised using the far-field value at the back of the wave, $\rho^{\dagger}\approx23.86 \text{ (a.u.)}$. 

\subsection{Bayesian inference of model parameters}

\begin{figure}[p!]
    \centering
    \includegraphics[width=\linewidth]{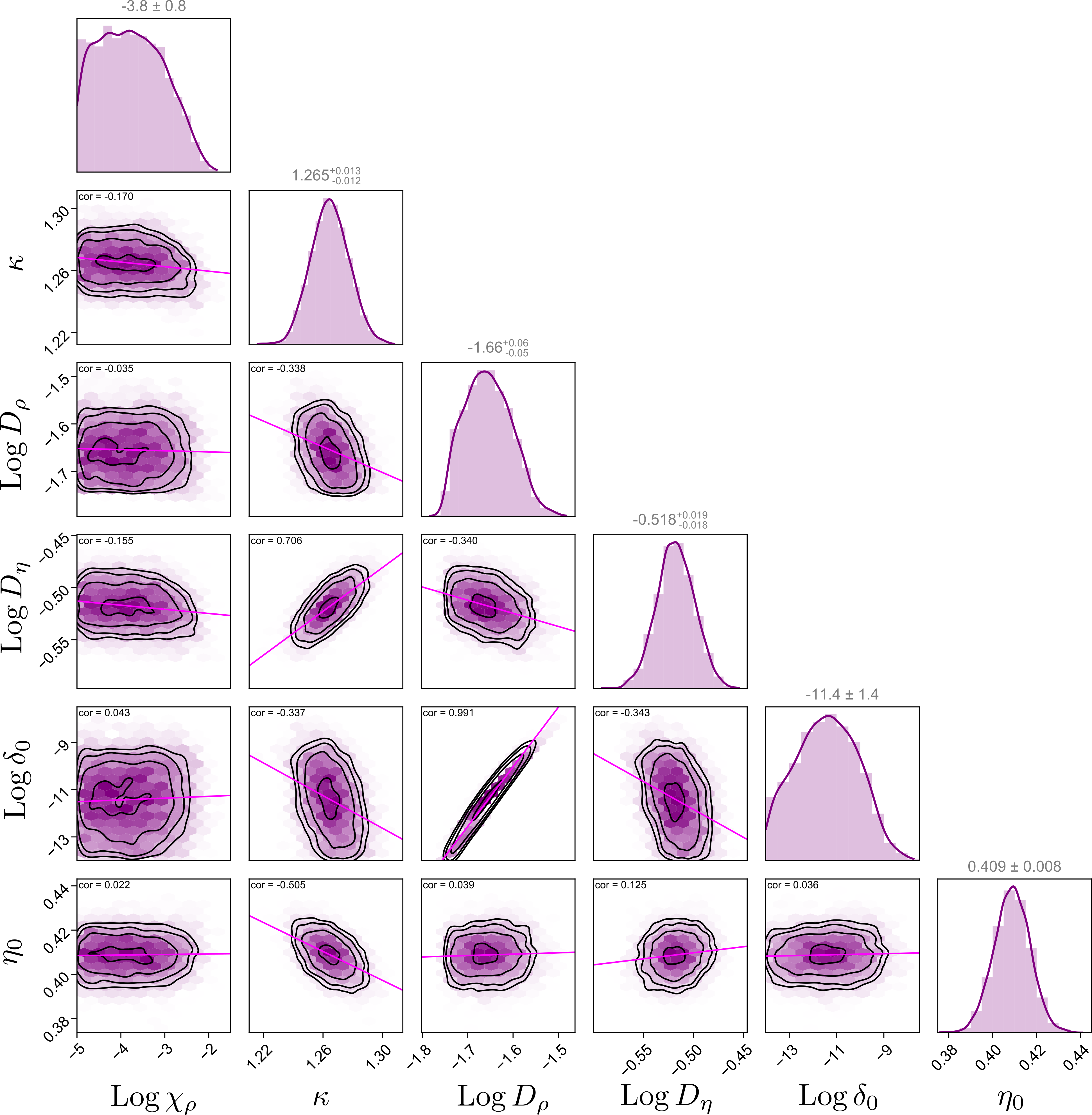}
    \caption{Results of the Hamiltonian Monte Carlo (HMC) algorithm for the CS travelling-wave system given by~\eqref{sys:fitting}. The plots along the diagonal represent the estimated \emph{marginal posterior distributions} for each model parameter. Below the diagonal, we plot the bivariate densities for every pair of parameters. Univariate posterior modes are indicated in Table~\ref{tab:estimates parameters}. Here, $\text{Log}:=\log_{10}$.}
    \label{fig:fitting results}
\end{figure}

While the travelling-wave system~\Cref{sys:fitting} is defined on an infinite domain, the experimental data only provide information on a finite portion of space. We equip~\Cref{sys:fitting} with the following set of initial conditions to fix the translational invariance of the original problem
\begin{equation}
    \rho(z_0)=1-\delta_0,\quad \eta(z_0)=\eta_0,\quad S(z_0)= 1-\alpha_S\delta_0\label{initial condition fitting}
\end{equation}
where $0\leq\delta_0\ll1$ and $\eta_0>0$ are unknown constants to be fitted together with the other model parameters, while $z_0$ is known and is the spatial location of the leftmost available experimental data. The additional constant $\alpha_S$ in~\eqref{initial condition fitting} is set as a function of the other model parameters
\begin{equation}
    \alpha_S=-D_\rho\left(\frac{1+\chi_\rho/2}{\chi_\rho}-\sqrt{\frac{(1+\chi_\rho/2)^2}{\chi_\rho^2}+\frac{1+\chi_\rho}{\chi_\rho D_\rho}}\right). \label{eq:expression alphaS}
\end{equation}
The expression~\eqref{eq:expression alphaS} is obtained by looking at the far-field behaviour of the solution near the back of the wave $(\rho,S)=(1,1)$; we are therefore assuming that the solution profile is close enough to the fixed point, which appears reasonable since the point $z_0$ is sufficiently far from the consumer profile. The full problem defined by~\eqref{sys:fitting}-\eqref{initial condition fitting} has therefore a total of 6 unknown parameters $\vec{\theta}=(\chi_\rho,D_\rho,D_\eta,\kappa,\delta_0,\eta_0)$. 

\par\vspace{2mm}
\noindent \textbf{Error Model.} The data set consists of measurements of the mean density for the two cell populations $D=\left\{(\rho^D(z_j),\eta^D(z_j))\right\}_{j\in\mathcal{J}}$ and the standard deviation $\Sigma^D=\left\{(\sigma_\rho(z_j),\sigma_\eta(z_j)\right\}$ estimated from aligning density profiles in the travelling wave profile for the last 10 frames of the video.
In order to connect experimental data and models, we assume that the observations $(\rho^D,\eta^D)$ are noisy versions of the model-predicted density $(\rho,\eta)$. For simplicity, and using a common approach in the literature, we assume that the observation errors are additive, independent, and normally distributed with variances $(\sigma^2_\rho,\sigma^2_\eta)$ obtained directly from the data. This leads to the specification of the following error model
\begin{equation}
    \varphi^{D}(z_j)= \varphi(z_j;\vec{\theta}) + \sigma^2_\varphi(z_j) \ \mathcal{N}(0,1),\quad \varphi\in\left\{\rho,\eta\right\},\label{eq:error model}
\end{equation}
where $\mathcal{N}$ indicates the \emph{Gaussian distribution}, $\varphi(\cdot;\vec{\theta})$ indicates the model predictions for $\varphi$ given the parameter values $\vec{\theta}$, and $\sigma_\varphi(z_j)$ are the experimentally measured standard deviation for the density profiles of the two populations. Based on the error model~\eqref{eq:error model}, we write the log-likelihood of observing the measured data given the parameter values $\vec{\theta}$ as
\begin{equation}
    \lik(\vec{\theta})\sim -\frac{1}{2}\sum_j \sum_{\varphi=\rho,\eta}\log(2\pi\sigma_\varphi(z_j))+\left(\frac{\varphi(z_j;\vec{\theta})-\varphi^D(z_j)}{\sigma_\varphi(z_j)}\right)^2.~\label{eq:likelihood}
\end{equation}
Pointwise estimates of model parameters can be obtained by maximising the likelihood function $\mathcal{L}_D$. We perform the likelihood optimization using the parameter inference package \texttt{Turing}~\cite{Turing}. In particular, we obtain a series of maximum likelihood estimates $\vec{\theta}_{ML}$ using \emph{Particle Swarm optimisation}, which explores the parameter space dynamically by evolving a population of candidate solutions. To generate the initial configurations for the particle population, we sample each parameter from an independent uniform distribution across a range of physically relevant parameter regimes
\begin{equation}
    \log_{10}\chi_\rho\sim \Unif{-4.5}{2}, \quad \log_{10} D_\rho\sim\Unif{-5}{2},\quad \log_{10} \frac{D_\eta}{\kappa-1}\sim\Unif{-2.5}{2},\quad \kappa\sim\Unif{1}{10},\quad\log_{10}\delta_0\sim\Unif{-14}{-1},\quad \rho_0\sim \Unif{0}{1}.\label{eq:uniformprior}
\end{equation}
Note that we use log-uniform sampling for those parameters whose values can span across multiple orders of magnitude. Furthermore, we do not fit $D_\eta$ directly; rather, we rescale it by the factor $(\kappa-1)$ to control the growth of the sensor cells distribution at the back of the wave and stabilise the integration of~\eqref{sys:fitting}. Since we aim to obtain a good-enough estimate of the optimal value of the model parameters, we limit the number of iterations for the \emph{Particle Swarm optimization} to 500. Repeating the procedure, we obtain consistent estimates for $\vec{\theta}_{ML}$ across repeats; we denote by $\hat{\vec{\theta}}\approx[10^{---4.5},10^{-1.603},10^{-0.528},1.262,10^{-10.15},0.405]$ the parameter set that yields the maximum likelihood estimates across repeats. We adopt \emph{Bayesian inference} to refine our point estimates of model parameters $\hat{\vec{\theta}}$ and their uncertainty. The aim is to obtain an approximation to the posterior distribution $P(\vec{\theta}|D)$, which can be written via \emph{Bayes Theorem} as
\begin{equation}
    P(\vec{\theta}|D)\sim \lik(\vec{\theta}) \pi(\vec{\theta}),
\end{equation} 
where the likelihood function $\lik$ is defined by~\eqref{eq:likelihood} and $\pi(\vec{\theta})$ is the prior, which captures information about the value of model parameters. Here, we again use an uninformative uniform prior~\eqref{eq:uniformprior} to avoid biases in the sampling of the posterior
\begin{equation}
    \log_{10}\chi_\rho\sim \Unif{-5}{0}, \quad \log_{10} D_\rho\sim\Unif{-3}{0},\quad \log_{10} \frac{D_\eta}{\kappa-1}\sim\Unif{-1}{1},\quad \kappa\sim\Unif{1}{2},\quad\log_{10}\delta_0\sim\Unif{-15}{-2},\quad \rho_0\sim \Unif{0}{1}.\label{eq:priorA}
\end{equation}%Here, we construct the prior assuming a Gaussian spread around the maximum-likelihood estimates $\hat{\vec{\theta}}$
%\begin{equation}
 %   \theta\sim\hat{\theta}\ (1+\ \mathcal{N}(0,0.25)), \quad \theta\in\left\{\log_{10}\chi_\rho,\log_{10}D_\rho,\log_{10}D_\eta,\kappa,\log_{10}\delta_0,\eta_0\right\}.
%\end{equation}
In order to infer the posterior distribution $P(\vec{\theta}|D)$, we use the \emph{No-U-Turn Sampler}~\cite{hoffman2014no}, which is an extension of Hamiltonian Monte Carlo (HMC). As such, it requires differentiation of the likelihood function $\lik$; this is estimated by automatic differentiation through the SciML ecosystem in Julia, which allows for Bayesian inference of differential equations via the combination of \texttt{Turing.jl} and \texttt{DifferentialEquations.jl} packages. We use the NUTS algorithm to generate 4 independent chains with a target acceptance rate of $0.75$. All chains are started from the same maximum-likelihood estimate $\hat{\vec{\theta}}$. Each chain is of length 3000, but we consider only the last 2000 iterations of each chain to construct the posterior distributions.

\subsection{Results calibration}\label{sec:calibration results} 
\begin{figure}[htb]
    \centering
    \includegraphics[width=\linewidth]{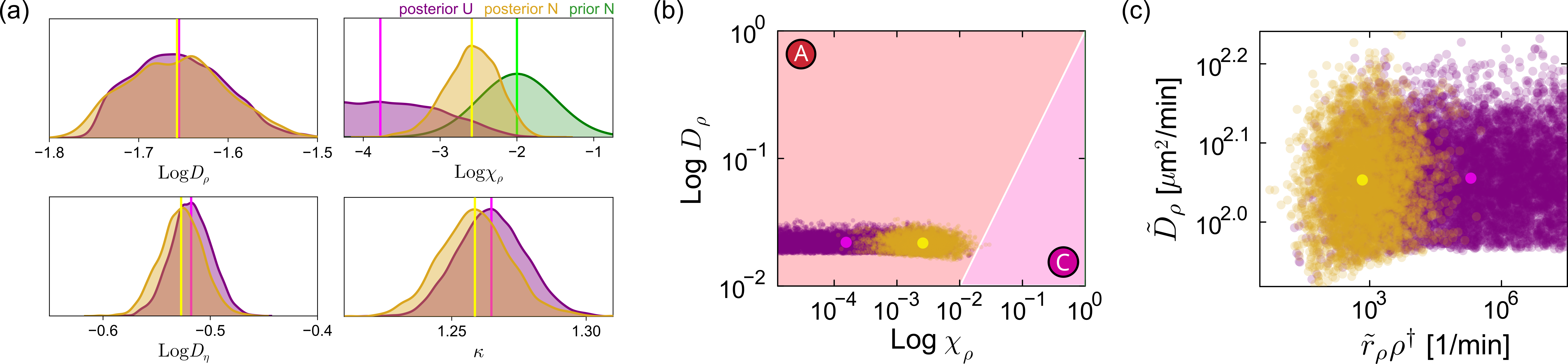}
    \begin{subfigure}{0.0\textwidth}      \captionlistentry{}
    \label{fig:additionalfittingA}\end{subfigure}
        \begin{subfigure}{0.0\textwidth}      \captionlistentry{}
    \label{fig:additionalfittingB}\end{subfigure}
        \begin{subfigure}{0.0\textwidth}      \captionlistentry{}
    \label{fig:additionalfittingC}\end{subfigure}
    \vspace{-5mm}
    \caption{Sensitivity of parameter inference to the prior distribution for $\chi_\rho$. (a) Comparison of the marginal posterior distributions of non-dimensional model parameters obtained using an uninformative uniform (U) -- purple curve -- and an informative Normal (N) --yellow curve -- prior for $\text{Log}\,\chi_\rho$. The normal prior for $\text{Log}\,\chi_\rho$ is shown in green. (b) Comparison of the joint posterior distribution of $\text{Log}\,\chi_\rho$ and $\text{Log}\, D_\rho$ obtained using the two prior choices. Shaded regions indicate the asymptotic regimes defined by the relative magnitude of $\chi_\rho$ and $D_\rho$ (\Cref{sec:asymptotic analysis migrating fronts} of the main text). The magenta and light yellow dots indicate the corresponding posterior median estimate. (c) Joint posterior distributions of the dimensional consumer diffusion coefficient $\tilde D_\rho$ and the net chemoattractant consumption rate at the rear of the front, $\tilde r_\rho\rho^\dagger$, with the latter obtained from the wave-speed relation~\eqref{eq:velocity}. Colours correspond to the prior choices in panel (a). Here, $\text{Log}:=\log_{10}$.}
    \label{fig:additionalfigurefitting}
\end{figure}

In~\Cref{fig:fitting results}, we show the marginal and bivariate marginal posterior distributions estimated via HMC as described above. The corresponding summary statistics for the univariate marginal posterior distribution, namely median and interquartile range, are given in~\Cref{tab:estimates parameters}. Figure 1a in the main text is obtained by sampling 1000 independent sets of parameters from the posterior distribution. Generally, we find that the univariate densities are unimodal and symmetric, allowing us to constrain most model parameters except for $\chi_\rho$. The distribution of $\chi_\rho$ spans several orders of magnitude, and it tends to a uniform distribution once $\chi_\rho$ is sufficiently small,$\log_{10}\chi_{\rho}\lessapprox-3$. While we can not precisely identify the value of $\chi_\rho$, we find significant statistical evidence that its value is small and that $\chi_\rho <D_\rho$ (purple area in~\Cref{fig:additionalfittingB}). We note this is in contrast to the choice of model parameters adopted in~\cite{Ucar2025} (see~\Cref{tab:estimates parameters}), where $\chi_\rho$ was chosen to be larger than the cell diffusivity based on qualitative (rather than quantitative) agreement with experimental observations. When considering the other model parameters that were obtained in~\cite{Ucar2025} either via indirect measurement or via a linear approximation of the wave front, we find that our estimates are in good agreement (see $\kappa$, $D_\rho$, and $D_\eta$ in Table~\ref{tab:estimates parameters}). 
\begin{table}[h]
    \centering
    \begin{tabular}{c|c c c c c c}
    \toprule[1.5pt]
        Parameter & $\kappa$ & $\log_{10}D_\rho$ & $\log_{10} D_\eta$ & $\log_{10}\chi_\rho$ & $\log_{10}\delta_0$ & $\eta_0$ \\
        \midrule
        Median & 1.2646&-1.6572&-0.5180 &-3.8089&-11.4081&0.4089\\[2pt]
        IQR & [1.256,1.273]& [-1.693,-1.619] & [-0.530,-0.505]&[-4.380,-3.213] &[-12.384,-10.448]& [0.404,0.414]\\
        Estimates in \cite{Ucar2025} & 1.26 & -1.155& -0.700 &-0.700& NA & NA\\
        \bottomrule[2pt]
    \end{tabular}
    \vspace{2mm}
    \caption{Posterior estimates of model parameters. For each parameter in~\eqref{sys:fitting}, we indicate the median and interquartile range (IQR) of the corresponding posterior marginal distribution. We also compare the obtained estimates with the parameters used by Ucar et~al. (values taken from Supplementary Information Table S1~\cite{Ucar2025}).}
    \label{tab:estimates parameters}
\end{table}

The uncertainty in the estimates of $\chi_\rho$ propagates to the net chemoattractant consumption rate at the rear of the advancing front, \emph{i.e.}, $\tilde{r}_\rho\rho^\dagger$, determined by~\eqref{eq:velocity} (purple area in~\Cref{fig:additionalfittingC}). In particular, since $\tilde{r}_\rho\rho^\dagger \propto\chi_{\rho}^{-2}$ when $\chi_\rho\ll1$, we find that consumption is fast, beyond realistic timescales (see~\Cref{fig:additionalfittingC}) -- with an effective bulk chemoattractant depletion rate of approximately $0.1$ [1/hr] estimated from dendritic-cell suspensions~\cite{alanko_ccr7_2023}). This discrepancy may partly reflect differences in cell density and set-up between the migration and depletion experiments. Alternatively, it may indicate that the minimal model compensates for mechanisms not represented explicitly in~\eqref{sys:fitting} by favouring very small chemotactic sensitivities and correspondingly large effective consumption rates.

To test whether the inferred asymptotic ordering depends on the weak identifiability of \(\chi_\rho\), we repeat the inference using an informative prior favouring larger values of \(\chi_\rho\). We centre the prior at \(\mu_\chi=-2\), thereby penalising the very small values favoured under the original uninformative prior. Specifically, we define the prior distributions of model parameters as
\begin{equation}
    \log_{10}\chi_\rho\sim \mathcal{N}(\mu_\chi,\sigma_\chi^2), \quad \log_{10} D_\rho\sim\Unif{-3}{0},\quad \log_{10} \frac{D_\eta}{\kappa-1}\sim\Unif{-1}{1},\quad \kappa\sim\Unif{1}{2},\quad\log_{10}\delta_0\sim\Unif{-15}{-2},\quad \rho_0\sim \Unif{0}{1},\label{eq:priorB}
\end{equation}
where $\sigma_\chi=0.25$. The comparison of the marginal posterior distributions of key model parameters for priors~\eqref{eq:priorA} (purple curves) and~\eqref{eq:priorB} (yellow curves) is illustrated in~\Cref{fig:additionalfittingA}. The posterior distributions of \(\kappa\), \(D_\rho\), and \(D_\eta\) are essentially unchanged. Although the informative prior shifts the posterior for \(\chi_\rho\) towards larger values, the posterior remains substantially shifted towards smaller values relative to the prior and predominantly satisfies \(\chi_\rho<D_\rho\) (\Cref{fig:additionalfittingB}). Consequently, we conclude that while the absolute value of \(\chi_\rho\) is prior-sensitive, its asymptotic ordering relative to \(D_\rho\) is robust. The loss of identifiability of \(\chi_\rho\) when this is sufficiently small is consistent with the asymptotic analysis in~\Cref{sec:asymptotic analysis migrating fronts}: once \(\chi_\rho\ll D_\rho\), the leading-order travelling-wave profile becomes insensitive to the precise value of \(\chi_\rho\). Thus, the spatial data identify the asymptotic regime rather than the absolute value of this parameter. Hence, we can still learn the physical balance that governs the collective migration of dendritic and T cells.

\section{Asymptotic analysis of the spatial structuring of the wave front}
\label{sec:asymptotic analysis migrating fronts}
In this section, we detail the derivation of the results presented in Section 3 of the main text, which rely on asymptotic analysis of~\eqref{sys:fitting}. Owing to the simplifying assumption that consumption arises solely from the consumer population, we can focus on the solution structure of the reduced system
\begin{subequations}
\begin{align}
    D_\rho\rho'&= \rho\left(S-1\right),\\
    \chi_\rho S'&=\rho(\chi_\rho+1)-S\left(S+\chi_\rho\right).\label{eq S asymptotics}
\end{align}\label{sys:asymptotics}%
\end{subequations}
Given the signal $S=S(z)$, the distribution of the sensor population can be explicitly obtained as
\begin{equation}
    \eta(z)=K_\eta\exp\left(\frac{1}{D_\eta}\int\kappa S-1\, dz\right),\label{eq:eta implicit S}
\end{equation}
where the constant $K_\eta>0$ is defined by imposing the total number of sensor cells. Hence, we start focusing on the analysis of~\eqref{sys:asymptotics} which defines the structure of the migrating front. Examples of solutions to the subsystem~\eqref{sys:asymptotics} for the values of the parameters listed in~\Cref{tab:Figure2 main text} are illustrated in Figure 2 of the main text. 

We start by considering the two distinguished limits of small- and large-$D_\rho$. This analysis corresponds to taking a shallow and sharp interface limit for the migrating front, respectively. We shall see that the analysis of these two limits reveals the presence of four distinguished sublimits, depending on the size of the chemotactic sensitivity $\chi_\rho$ (Figure 2 in the main text). For these four sublimits A-D, we derive an approximate asymptotic solution.
\begin{table}[ht!]
    \centering
    \begin{tabular}{c||c c c c}
    \toprule
         Regime &  A &B&C&D\\
         \midrule
         $D_\rho$& 25 & 25& 0.0001 & 0.1\\
         \midrule
         $\chi_\rho$& 0.01& 100 &0.001 &100\\
         \bottomrule
    \end{tabular}
    \vspace{2mm}
    \caption{List of the model parameters used to generate Figures 2 and 3 in the main text.}
    \label{tab:Figure2 main text}
\end{table}

\subsection{Diffusive fronts: $D_\rho\gg1$}
\begin{figure}[htb!]
        \centering
        \begin{subfigure}{0\textwidth}        
            \captionlistentry{}
            \label{fig: asymptotic smooth A}
        \end{subfigure}
        \begin{subfigure}{\textwidth}        
            \captionlistentry{}
        \includegraphics[width=\linewidth]{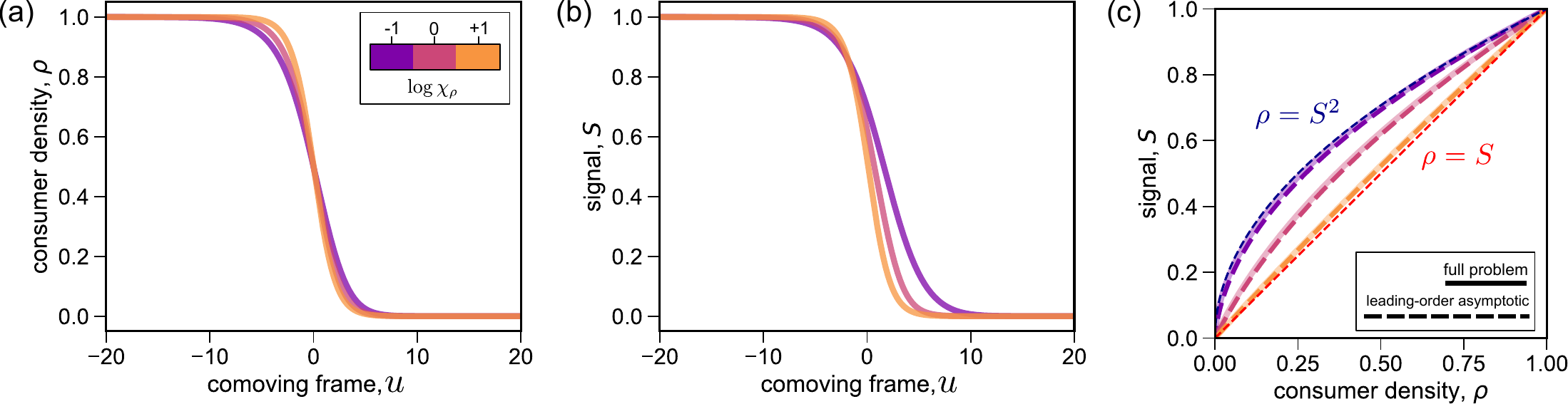}
             \label{fig: asymptotic smooth B}
        \end{subfigure}
                \begin{subfigure}{0\textwidth}        
            \captionlistentry{}
            \label{fig: asymptotic smooth C}
        \end{subfigure}
        \caption{(a)-(b) Spatial profile of the solution to the TW problem~\eqref{sys:asymptotics}. Different colours indicate distinct values of $\chi_\rho$, while the parameter $D_\rho$ is fixed to $D_\rho=10$. (c) Solution to~\eqref{sys:asymptotics} in the $(\rho,S)$ phase plane (see full curves). The dashed line corresponds to the prediction of the leading-order asymptotic for the corresponding value of $\chi_\rho$~\eqref{limit 1 relation rho S}. The dashed blue and red lines correspond to the predicted asymptotic behaviour when $\chi_\rho\ll1$~\eqref{explicit solution small chi limit} and $\chi_\rho\gg1$~\eqref{explicit solution large chi limit}, respectively.}
        \label{fig:asymptotic general limit A}
    \end{figure}

  To resolve variation in the consumer cell density in the limit of large-$D_\rho$, we first rescale the spatial dimension by $D_\rho$
    \[Z=\frac{z}{D_\rho},\]
    to yield
    \begin{subequations}
    \begin{align}
        \rho'&= \rho\left(S-1\right),\\
         \frac{\chi_\rho}{D_\rho}S'&=\rho(\chi_\rho+1)-S\left(S+\chi_\rho\right),\label{eq: S diffusive interface}
    \end{align}\label{sys: asymptotic smooth}%
    \end{subequations}
    with far-field conditions~\eqref{eq: far-field}.
    While the expression for the consumer cell density $\rho$ contains only $\mathcal{O}(1)$ terms, the dominant physical balance shaping the signal $S$ depends on the size of $\chi_\rho$. Figure~\ref{fig:asymptotic general limit A} illustrates how solutions to~\eqref{sys: asymptotic smooth} in the large-$D_\rho$ limit depend on the size of $\chi_\rho$. The latter affects both the spatial distribution for the consumer $\rho$ and signal $S$, and their relationship (\Cref{fig: asymptotic smooth C}). However, independently of $\chi_\rho$, the leading-order behaviour for $S$ will always be defined by equilibrating the right-hand side of~\eqref{eq: S diffusive interface}, while the spatial variation in $S$ remains negligible.
     We can resolve the global structure of the solution simply by looking at the leading-order behaviour of the solution based on a standard power series expansion (or \emph{Poincar\'e expansion})
        \begin{equation}
            \rho=\rho_0+\mathcal{O}(D^{-1}_\rho),\quad S= S_0+\mathcal{O}(D^{-1}_\rho).\label{eq:poincare' expansion}
        \end{equation}
        Substituting the above ansatz into~\eqref{sys:fitting}, we find that the leading-order problem for $\rho_0$ and $S_0$ is
        \begin{subequations}
            \begin{align}
             \rho_0'&=\rho_0(S_0-1),\\
             \rho_0&=\frac{S_0(\chi_\rho+S_0)}{1+\chi_\rho},\label{limit 1 relation rho S}
            \end{align}\label{reduced ODE TW problem limitA}%
        \end{subequations}
        which implies that, at leading order, the signal $S$ is fully determined by the distribution of the consumer populations (or vice versa). When comparing~\eqref{limit 1 relation rho S} with the solution of the full problem for relatively large values of $\sigma_\rho$, we find good agreement (see panel (c) in~\Cref{fig:asymptotic general limit A}). We can use the two expressions above to find a closed-form expression for the distribution of $S_0$ in the travelling-wave frame 
        \begin{equation}
        S_0'=\frac{S_0(\chi_\rho+S_0)(S_0-1)}{\chi_\rho+2S_0}.\label{limitI:dS}
        \end{equation}
        While we cannot solve~\eqref{limitI:dS} explicitly, we can still apply separation of variables to~\eqref{limitI:dS} to yield the following implicit expression for $S_0(Z)$
        \begin{equation}
            \frac{(1-S_0)^{\chi_\rho+2}}{S_0^{\chi_\rho+1}(\chi_\rho+S_0)}=\front{K}_s e^{(\chi_\rho+1)Z},\label{limitI:S}
        \end{equation}
        which satisfies both boundary conditions at the front and back of the wave. The additional degree of freedom $\front{K}_s$ is defined by fixing the translational symmetry of the solution. 
        \subsubsection{Sublimits A and B}
        While we do not have an explicit solution for $\rho$ or $S$, we can still make some relevant progress in specific limits of $\chi_\rho$. In particular, when taking the sublimit of $\chi_\rho\rightarrow\infty$ we find:
        \begin{subequations}
            \begin{align}
                \rho\sim S\sim\frac{1}{\front{K}_se^{z/D_\rho}+1}.
        \end{align}
        We then obtain the corresponding spatial distribution for the sensor population, substituting the expression above into~\eqref{eq:eta implicit S}
        \begin{align}
             \eta\sim K_\eta \left[\frac{e^{z(\kappa-1)/D_\rho}}{(\bar{K}_se^{z/D_\rho}+1)^{\kappa}}\right]^{1/\nu}.
        \end{align}\label{explicit solution large chi limit}        
        \end{subequations}
        In contrast, in the sublimit $\chi_\rho\rightarrow0$, we find
        \begin{equation}
            S\sim \frac{1}{\front{K}_se^{z/2D_\rho}+1}, \quad \rho\sim S^2\sim\left(\frac{1}{\front{K}_se^{z/2D_\rho}+1}\right)^2,\quad \eta\sim{K}_\eta \left[\frac{e^{z(\kappa-1)/D_\rho}}{(\bar{K}_se^{z/2D_\rho}+1)^{2\kappa}}\right]^{1/\nu}.\label{explicit solution small chi limit}
        \end{equation}
    We find good agreement between the asymptotic solutions~\eqref{explicit solution large chi limit}-\eqref{explicit solution small chi limit} with the solution to the full nonlinear problem~\eqref{sys: asymptotic smooth} (see Figure 2b in the main text and~\Cref{fig:asymptotic general limit A}). This suggests that the leading-order approximation is a good approximation to the solution and captures the main features of the solutions.
    
\subsection{Sharp front: $D_\rho\ll 1$}
\begin{figure}
    \centering
    \begin{subfigure}{0.0\textwidth}
        \captionlistentry{}
        \label{fig:asymptotic general limit B panel a}
    \end{subfigure}
        \begin{subfigure}{\textwidth}
        \includegraphics[width=\linewidth]{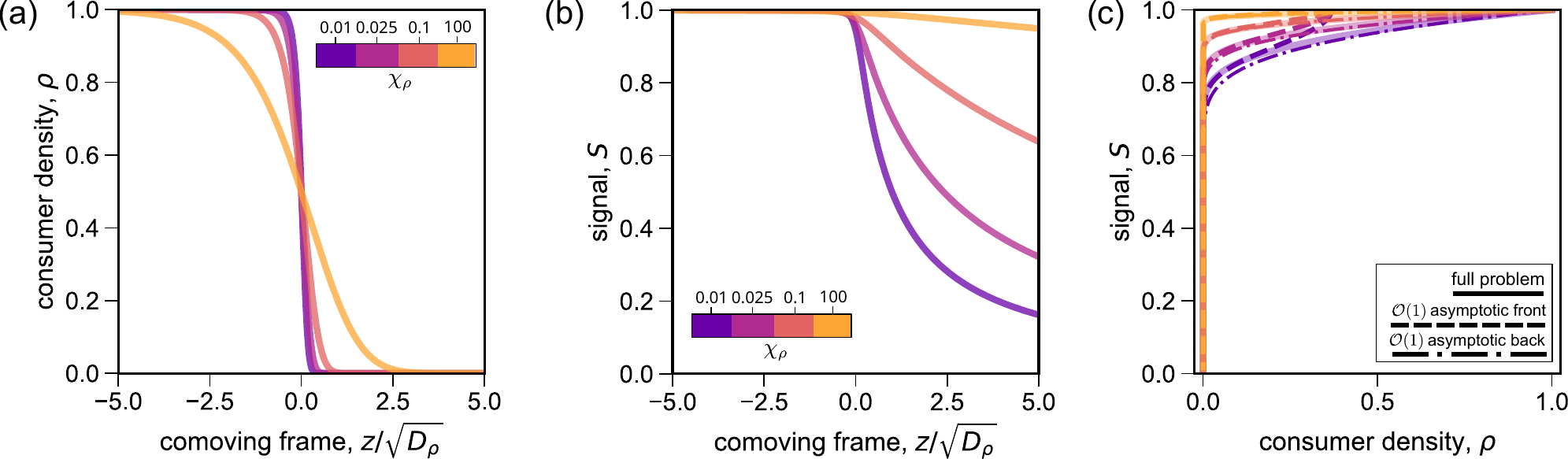}
          \captionlistentry{}
        \label{fig:asymptotic general limit B panel b}
        \end{subfigure}
    \begin{subfigure}{0.0\textwidth}
        \captionlistentry{}
        \label{fig:asymptotic general limit B panel c}
    \end{subfigure}
    \caption{(a)-(b) Spatial profile of the solution to the TW problem~\eqref{sys:asymptotics}. Different colours indicate distinct values of $\chi_\rho$, while the consumer diffusion coefficient is fixed to $D_\rho=0.0001$. (c) Solution to~\eqref{sys:asymptotics} in the $(\rho,S)$ phase plane (see full curves). The dashed line corresponds to the prediction of the leading-order asymptotic relationship between $\rho$ and $S$ ahead of the consumer interface~\eqref{eq:front final leading-order} and along the interface~\eqref{eq:first integral}.}
        \label{fig:asymptotic general limit B}
\end{figure}

     We write $D_\rho=\varepsilon^2$ and take the limit $\varepsilon\rightarrow 0$. In this case, there is no global dominant balance along the front, and we have to resolve two distinct asymptotic regions. In the outer region, corresponding to the front of the wave where $\rho\ll1$, chemoattractant consumption is negligible and the signal $S$ is shaped by transport. In contrast, in the inner region corresponding to the consumer front, the consumer cell density sharply transitions to being an $\mathcal{O}(1)$-quantity and the signal variable $S\sim 1$ (see Figures~\ref{fig:asymptotic general limit B panel a}-\ref{fig:asymptotic general limit B panel b}).  We adopt a matched asymptotic expansion approach to connect these two distinct regions. Here, the asymptotic matching requires tracking the exponentially small corrections, while standard \emph{Poincaré} expansions in powers of $\varepsilon$, as in~\eqref{eq:poincare' expansion}, fail.
\subsubsection{Front of the wave: $z\gg 1$}
We start by considering the behaviour of the solution in the far-field ahead of the migrating front ($z\gg 1$). Using the $\varepsilon$ notation, we rewrite~\eqref{sys:asymptotics} as
\begin{subequations}
    \begin{align}
        \varepsilon^2\rho'&= \rho(S-1),\label{eq rho limit B}\\
        S'&=\frac{(1+\chi_\rho)\rho-S(S+\chi_\rho)}{\chi_\rho},
    \end{align}
     with the far-field condition
     \begin{align}
         \rho,S\rightarrow 0, \quad z\rightarrow \infty.\label{eq: BC front limit B}
     \end{align}\label{sys limit B}
\end{subequations}
We note that the trivial solution
\[(\rho,S)\equiv(0,0),\]
is an exact solution for the system~\eqref{sys limit B}. As a result, deviations from the trivial solution will involve exponentially small corrections. In order to balance the terms in~\eqref{eq rho limit B}, we use the following expansion around the trivial state
\begin{equation}
(\rho,S)\sim\left(\exp\left(-\frac{ f_0}{\varepsilon^2}\right),S_0\right),\label{eq:front asymptotic expansion}
\end{equation}
which yields the following leading-order problem for $f_0$ and $S_0$
\begin{subequations}
\begin{align}
    f_0'&=1-S_0,\\
    S_0'&=-S_0\left(1+\frac{S_0}{\chi_\rho}\right),
\end{align}    
with boundary conditions
\begin{align}
 f_0(z)\rightarrow\infty, \quad S_0(z)\rightarrow0 \quad \text{ as }z\rightarrow \infty.
\end{align}\label{eq:leading-order front}
\end{subequations}
We can integrate~\eqref{eq:leading-order front} exactly to obtain
\[f_0=\front{f}_0-\chi_\rho\ln(\front{K}_s-e^{-z})+z, \quad S_0=\frac{\chi_\rho}{\front{K}_s e^{ z}-1},\]
resulting in the following leading-order behaviour for the original variables $S$ and $\rho$
\begin{equation}
S\sim\frac{\chi_\rho}{\front{K}_s e^{ z}-1}, \quad \rho\sim \front{K}_\rho e^{-\frac{(1+\chi_\rho) z}{\varepsilon^2}}\left(\front{K}_se^{z}-1\right)^{\frac{\chi_\rho}{\varepsilon^2}}.\label{eq:front leading-order solution}
\end{equation}
In~\eqref{eq:front leading-order solution}, there are two unknown constants $\front{K}_s$ and $\front{K}_\rho$, which are not set by the far-field behaviour~\eqref{eq: BC front limit B}. %and $\front{K}_\eta$. 
One degree of freedom results from the fact that the original travelling wave problem~\eqref{sys:fitting} is translationally invariant, and can therefore be fixed arbitrarily. Without loss of generality, we set $\front{K}_s=\chi_\rho+1$ leading to 
\begin{equation}
    S\sim\frac{\chi_\rho}{(\chi_\rho+1)e^{z}-1}, \quad \rho\sim \front{K}_\rho e^{-\frac{z}{\varepsilon^2}}\left(\chi_\rho+1-e^{-z}\right)^{\frac{\chi_\rho}{\varepsilon^2}}.\label{eq:front final leading-order}
\end{equation}
The remaining degree of freedom, $\front{K}_\rho$ is fixed by the matching of~\eqref{eq:front final leading-order} to the behaviour of the solution in Region II of the domain. We note that~\eqref{eq:front final leading-order} is not valid for $z<0$ since the signal goes past its maximal value, \emph{i.e.}, $S_0>1$. Indeed, when $S\sim1$, the balance of the terms in~\eqref{eq rho limit B} changes, and the expansion~\eqref{eq:front leading-order solution} breaks down. In particular, we find that at $z_0=0$ a boundary layer forms, where the cell density $\rho$ rapidly decreases from $\rho=\mathcal{O}(1)$ to $\rho$ being exponentially small. In Figure~\ref{fig:asymptotic general limit B panel c}, we compare the asymptotic solution~\eqref{eq:front final leading-order} to the solution of~\eqref{sys: asymptotic smooth} in the $(\rho,S)$-plane. We find good agreement across values of $\chi_\rho\gg D_\rho$.

\subsubsection{Boundary layer analysis: $z=\mathcal{O}(\varepsilon)$}
In order to match the behaviour at the front of the wave~\eqref{eq:front final leading-order} to the far-field behaviour~\eqref{eq: far-field} as $z\to\infty$, we have to resolve the boundary layer near $z=0$. With the benefit of hindsight, we resolve this region by imposing the scalings
\begin{equation}
    S=1-\varepsilon \sqrt{\frac{\chi_\rho+1}{\chi_\rho}}Y, \quad z=\varepsilon\sqrt{\frac{\chi_\rho}{\chi_\rho+1}} u.\label{eq:scaling sharp inner}
\end{equation}
We note that the boundary layer being located at $z=0$ is not generic, and follows from how we have fixed the translational invariance at the front of the wave to ensure that $S\sim 1$ near the origin. 
Substituting~\eqref{eq:scaling sharp inner} into~\eqref{sys:asymptotics}, we find
\begin{subequations}
    \begin{align}
    \rho'&=-\rho Y,\\
     (\chi_\rho+1) Y'&=-(1+\chi_\rho)\left(\rho-1+\varepsilon \sqrt{\frac{\chi_\rho+1}{\chi_\rho}} Y\right)-\varepsilon \sqrt{\frac{\chi_\rho+1}{\chi_\rho}} Y\left(1-\varepsilon  \sqrt{\frac{\chi_\rho+1}{\chi_\rho}}Y\right),
    \end{align}\label{sys:blayer}%
    with the far-field condition $(\rho,Y)\to(1,0)$ as $u\to-\infty$.
\end{subequations}
Substituting~\eqref{eq:scaling sharp inner} into~\eqref{eq:front final leading-order}, we obtain the conditions to match the behaviour in the boundary layer to the front of the wave:
\begin{align}
    Y\sim u,\quad \rho\sim \front{K}_\rho \chi_\rho^{\chi_\rho/\varepsilon^2}e^{-\frac{u^2}{2}},\quad u\rightarrow +\infty.
\end{align}
Next, we expand all independent variables in the inner region using a standard power series expansion
\[\rho\sim\rho_0+\varepsilon\rho_1, \quad Y\sim Y_0+\varepsilon Y_1.\]
and find that~\eqref{sys:blayer} at the leading-order simplifies to
\begin{subequations}
    \begin{align}
    \rho_0'&= -Y_0\rho_0,\label{eq:rho inner}\\[2pt]
    Y_0'&=-(\rho_0-1).
\end{align}
with a far-field condition
\begin{align}
    (\rho_0,Y_0)\to(1,0)\quad u\rightarrow -\infty,\quad (\rho_0,Y_0)\rightarrow (0,u) \quad u\rightarrow\infty \label{eq:blayer leading-order far-field}
\end{align}\label{sys:blayer leading-order}
\end{subequations}
Importantly,~\eqref{sys:blayer leading-order} corresponds to the leading-order behaviour of the problem only if $\varepsilon\ll\sqrt{\chi_\rho}$. The regime $\chi_\rho=\mathcal{O}(\varepsilon^2)$ corresponds to a distinguished limit for the system~\eqref{sys:asymptotics} which delimits the region of parameter space where the consumer front is smooth (A and B in Figure 2 of the main text) and sharp (C and D in Figure 2 of the main text). While we can not solve~\eqref{sys:blayer leading-order} explicitly, we find that the leading-order problem has the following first integral
\begin{equation}
\frac{Y_0^2}{2}=\rho_0-1-\ln\rho_0, \label{eq:first integral}
\end{equation}
obtained after imposing the far-field conditions $\rho_0,Y_0\rightarrow (1,0)$ as $u\rightarrow-\infty$. Further, matching to the far-field condition at the front of the wave ($u\to\infty$) allows us to find the unknown constant
\[\front{K}_\rho=\frac{1}{e\chi_\rho^{\chi_\rho/\varepsilon^2}},\]
which then fully determines the leading-order behaviour of the solution in the front of the wave.

\begin{figure}
    \centering
    \includegraphics[width=0.75\linewidth]{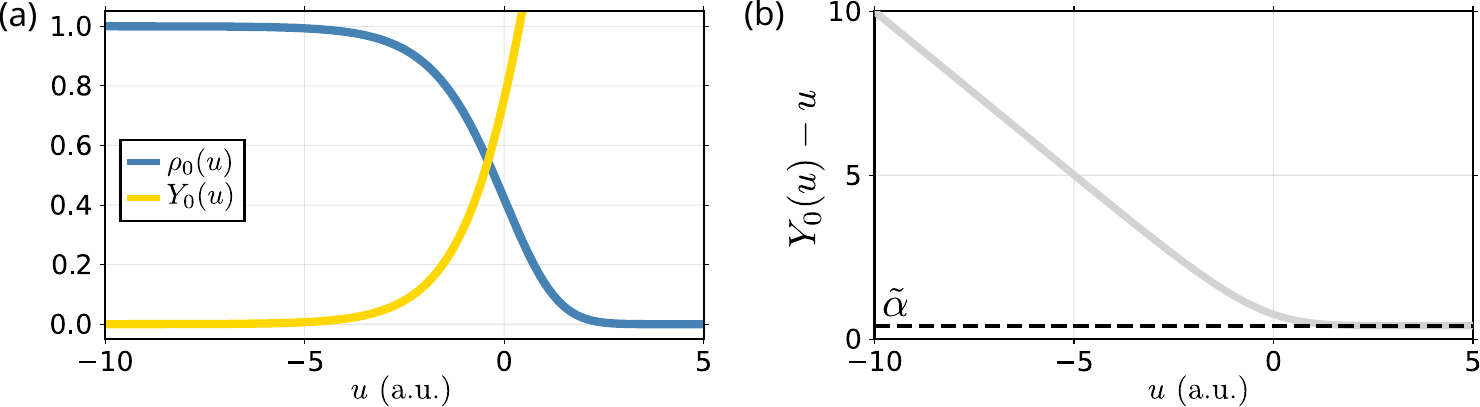}
    \caption{(a) Solution of~\eqref{sys:single ODE BL leading-order} via shooting (see main text for details). (b) Estimation of the matching constant $\tilde{\alpha}$ in~\eqref{eq:far field rho BL}.}
    \label{fig:resolving boundary layer}
\end{figure}

\paragraph{Numerical approximation of the cell density in the boundary layer.} We use~\eqref{eq:first integral} to express $Y_0$ as a function of $\rho_0$
\begin{equation}
    Y_0(\rho_0)=\sqrt{2(\rho_0-1-\ln\rho_0)},
\end{equation}
and reduce~\eqref{sys:blayer} to a single ODE for the variable $\rho$
\begin{subequations}
    \begin{align}
    \rho_0'=-\sqrt{2}\rho_0\sqrt{\rho_0-1-\ln\rho_0},
\end{align}
with the far-field condition
\begin{align}
    \rho_0\sim 1-A\exp(u),\quad u\rightarrow -\infty,\label{eq:rhofarfield_boundary_layer}
\end{align}\label{sys:single ODE BL leading-order}
\end{subequations}%
where $A>0$ is an (as-of-yet) unknown constant associated with the translational invariance of the wave, which is set by consistently matching the behaviour of $\rho_0$ in the far-field $u\to\infty$~\eqref{eq:blayer leading-order far-field}. To do this, we approximate the solution to~\eqref{sys:single ODE BL leading-order} numerically using a shooting technique. Specifically, without loss of generality, we fix $\rho_0(\bar{u})=1-\exp(\bar{u})$ with $\bar{u}\ll1$. Starting from $\bar{u}$, we integrate~\eqref{sys:single ODE BL leading-order} forward in space using the \texttt{solve} functionality implemented in \texttt{DifferentialEquation.jl} in \texttt{Julia}~\cite{rackauckas2017differentialequations}. We find that for $u\gg1$, the behaviour of the numerical solution satisfies:
\begin{equation}
    Y_0(\rho_0)-u=\back{\alpha}, \quad u\gg1~\label{eq:far field rho BL}
\end{equation}%
where the constant $\back{u}$ is dependent on the choice of $\bar{u}$. For example, setting $\bar{u}=-10$, we estimate $\back{\alpha}=0.412$ (\Cref{fig:resolving boundary layer}). To obtain the solution that satisfies the far-field behaviour, we simply shift the spatial variable by a factor of $-\back{u}$, implying that the value of the constant $A$ that matches to the front of the wave is $A=e^{\front{\alpha}}$.

As shown in Figure 2(b) of the main text, the obtained solution is in good agreement with the solution to the full problem in the small-$D_\rho$ regime, provided $\sqrt{D_\rho/\chi_\rho}\ll1$. 
\subsubsection{Composite solution}
Having obtained an explicit, closed-form solution for the solution in the front and back of the wave, we find that, in the singular limit $\varepsilon\to 0$, where the thickness of the boundary layer shrinks to zero, the leading-order behaviour for the variables $S$ and $\rho$ are:
\begin{equation}
    S\sim\begin{cases}
    1,&\quad z<0,\\
    \dfrac{\chi_\rho}{(1+\chi_\rho)e^z-1}, &\quad z>0,
    \end{cases}\qquad 
       \rho\sim\begin{cases}
    \rho_0\left(\dfrac{z(1+\chi_\rho)}{\varepsilon\sqrt{\chi_\rho}}\right),&\quad z<0,\\
    \dfrac{e^{-z/\varepsilon^2}}{e}\left(\dfrac{\chi_\rho+1-e^{-z}}{\chi_\rho}\right)^{\chi_\rho/\varepsilon^2}, &\quad z>0.
    \end{cases} \label{eq: leading-order sharp interface}
\end{equation}
Substituting~\eqref{eq:leading-order front} into~\eqref{eq:eta implicit S} allows us to compute also the spatial distribution of the sensor population:
\begin{equation}
  \eta\sim K_\eta\begin{cases}
     e^{(\kappa-1)z/D_\eta},&\quad z<0,\\
    e^{-z/D_\eta}\left(\frac{\chi_\rho+1-e^{-z}}{\chi_\rho}\right)^{\chi_\rho\kappa/D_\eta} &\quad z>0.\end{cases}
\end{equation}
\subsubsection{Sublimits C and D} 
We again leverage the additional parameter $\chi_\rho$ to simplify the expression obtained for the small-$D_\rho$ limit~\eqref{eq: leading-order sharp interface}.
First, we consider the short-signal case, when $D_\rho\ll\chi_\rho\ll1$ (see Figure~\ref{fig:asymptotic general limit B panel b}). Then the effective size of the interface is $\tilde{\varepsilon}=\sqrt{D_\rho\chi_\rho}$. To correctly balance the terms in~\eqref{eq S asymptotics}, we first rescale the spatial dimension $z=\tilde{\varepsilon}Z$ in the outer solution and expand the solution under the assumption of $\tilde{\varepsilon}\ll1$. We then rescale back to the $\mathcal{O}(1)$-spatial variable to obtain
% \begin{equation}
%     S\sim\begin{cases}
%     1,&\quad z<0,\\
%    \frac{1}{1+z/\chi_\rho}, &\quad z>0,
%     \end{cases}\qquad 
%        \rho\sim\begin{cases}
%     \rho_0(z/\tilde{\varepsilon}),&\quad z<0,\\[2pt]
%     \exp\left[-\dfrac{z^2}{2\tilde{\varepsilon}^2}-1\right], &\quad z>0,    \end{cases}
%  \qquad
%     \eta\sim K_\eta\begin{cases}
%      e^{(\kappa-1)z/D_\eta},&\quad z<0,\\
%     e^{-z/D_\eta}\left(1+\dfrac{z}{\chi_\rho}\right)^{\chi_\rho\kappa/D_\eta}&\quad z>0,%\exp\left[-\dfrac{\kappa(z-z_\eta)^2}{2\chi_\rho D_\eta}+\dfrac{\kappa z_\eta^2}{2\chi_\rho D_\eta}\right]&\quad z>0,
%     \end{cases} \end{equation}
\begin{align}
(S,\rho,\eta)\sim
\begin{cases}
\left(
1,\,
\rho_0(z/\tilde{\varepsilon}),\,
K_\eta e^{(\kappa-1)z/D_\eta}
\right), & z<0,\\[4pt]
\left(
\dfrac{1}{1+z/\chi_\rho},\,
\exp\left[-\dfrac{z^2}{2\tilde{\varepsilon}^2}-1\right],\,
K_\eta e^{-z/D_\eta}\left(1+\dfrac{z}{\chi_\rho}\right)^{\chi_\rho\kappa/D_\eta}
\right), & z>0.
\end{cases}
\end{align}
    where $\tilde{\rho}_0$ is the solution to~\eqref{sys:single ODE BL leading-order}. 

    The contrasting case of long-range chemical signalling corresponds to the limit $\chi_\rho\gg1$ (see Figure~\ref{fig:asymptotic general limit B panel b}). In this regime, the interfacial thickness is uniquely dependent on the consumer cell diffusivity $\varepsilon=\sqrt{D_\rho}$ and the approximate solution can be readily obtained:
 %    \begin{equation}
 %    S\sim\begin{cases}
 %    1,&\quad z<0,\\
 %   e^{-z}, &\quad z>0,
 %    \end{cases}\qquad 
 %       \rho\sim\begin{cases}
 %    \rho_0(z/\varepsilon),&\quad z<0,\\[2pt]
 %    \exp\left[-\dfrac{z-1+e^{-z}}{\varepsilon^2}-1\right], &\quad z>0,    \end{cases}
 % \quad
 %    \eta\sim K_\eta\begin{cases}
 %     e^{(\kappa-1)z/D_\eta},&\quad z<0,\\
 %    \exp\left[-\dfrac{z-\kappa(1-e^{-z})}{D_\rho}\right]&\quad z>0,
 %    \end{cases} \end{equation}
 \begin{align}
(S,\rho,\eta)\sim
\begin{cases}
\left(
1,\,
\rho_0(z/\varepsilon),\,
K_\eta e^{(\kappa-1)z/D_\eta}
\right), & z<0,\\[4pt]
\left(
e^{-z},\,
\exp\left[-\dfrac{z-1+e^{-z}}{\varepsilon^2}-1\right],\,
K_\eta\exp\left[-\dfrac{z-\kappa(1-e^{-z})}{D_\eta}\right]
\right), & z>0.
\end{cases}
\end{align}
    which implies that the maximum for the sensor populations is now located at $z\sim\ln\kappa$. Hence, for the same value of $\kappa$, the sensor cells are more likely to depart from the consumer in the case of long- compared to short-range signalling.
    
\section{Asymptotic analysis of sensor-consumer cell mixing}
Here, we leverage the asymptotic results from~\Cref{sec:asymptotic analysis migrating fronts} to compute the spatial distribution of sensor cells $\eta$ defined by~\eqref{eq:eta implicit S} and obtain analytical estimates of the spatial co-localisation of the consumer and sensor cells within the advancing front. For each sublimit A-D, we obtain a phase diagram that, depending on the properties of the sensor population, distinguishes distinct spatial structuring of the migrating front. To estimate the level of cell mixing, we introduce two metrics $\mu$ and $\sigma^2$ to quantitatively study the co-localisation of the two populations
\begin{equation}
\bar{\mu}=\alpha^{-1}\int_{0}^1 \eta  \rho d\rho,\quad \sigma^2=\alpha^{-1}\int_{0}^1 (\rho-\bar{\mu})^2\eta d\rho,
\quad \text{where } \alpha=\int_{0}^1 \eta d\rho, \label{eq: definition colocalisation metrics}
\end{equation}
where we treat $\eta$ as a function of the consumer cell density $\rho$ rather than space; note that since $\rho(z)$ is monotonically decreasing, the composite function $\eta(\rho)=\eta(z^{-1}(\rho))$ is well-defined. Since the mass of sensor cells is finite, the metrics $\mu$ and $\sigma^2$ are both well-defined, and they are independent of the choice of the spatial parametrisation. Our spatial colocalisation metrics~\eqref{eq: definition colocalisation metrics} differ from the one adopted in~\cite{Ucar2025}. In this previous work, it has been shown how optimal migration strategies can be achieved as a trade-off between the sensor population chemotaxis speed and their ability to remain mixed with the consumer population. The authors use a metric of mixing based on the Jensen-Shannon divergence to measure the difference in the distribution of consumer and sensor populations. However, this is not well-defined when studying the travelling wave regime, as the mass of the consumer is not finite.  Our mixing metric allows for analytical tractability while capturing the key mechanisms of interest. 

\subsection{Sublimit A: diffusive interface and long-range signalling}
    We start by rewriting $\eta$ as a function of the signal variable $S$
    \begin{equation}
    \eta(S)\sim K_\eta \,[S^{2}(1-S)^{2(\kappa-1)}]^{1/\nu},
    \end{equation}
    from which we deduce that the sensor population distribution relative to the signal follows a known probability distribution, $\mathcal{X}\sim$ \emph{Beta}$(\alpha,\beta)$ -- up to a normalisation constant -- with shape parameter $\alpha=2/\nu+1$ and $\beta=2(\kappa-1)/\nu+1$, where $\nu=D_\eta/D_\rho$. Using the fact that $\rho\sim S^2$ in this regime, we can rewrite the colocalisation metrics $\mu$ and $\sigma$~\Cref{eq: definition colocalisation metrics} in terms of moments of \emph{Beta}$(\alpha,\beta)$ for which we have explicit expressions in terms of the shape parameters $\alpha$ and $\beta$
\begin{subequations}
\begin{flalign}
 \bar{\mu}&=\frac{\moment{3}}{\moment{}}=\frac{(\alpha+1)(\alpha+2)}{(\alpha+\beta+2)(\alpha+\beta+1)}=\frac{2(1+\nu)(2+3\nu)}{(2\kappa+3\nu)(2\kappa+4\nu)},\\
 \sigma^2&=\frac{\moment{5}}{\moment{}}-\bar{\mu}_\eta^2=\frac{(\alpha+1)(\alpha+2)(\alpha+3)(\alpha+4)}{(\alpha+\beta+1)(\alpha+\beta+2)(\alpha+\beta+3)(\alpha+\beta+4)}-\bar{\mu}^2\quad\notag\\[2pt] \textcolor{white}{\sigma^2}&\textcolor{white}{=\frac{\moment{5}}{\moment{}}-\bar{\mu}_\eta^2}\  =\frac{\nu(\nu+1)(3\nu+2)(2\kappa+\nu-2)\,\big(7\kappa\nu+4\kappa+15\nu^{2}+9\nu\big)}
{(\kappa+2\nu)^{2}(\kappa+3\nu)(2\kappa+3\nu)^{2}(2\kappa+5\nu)}
\end{flalign}\label{expression mixing metrics small chi limit}%
\end{subequations}%
In Figure 1c of the main text, we illustrate how $\bar{\mu}$ and $\sigma$ change as a function of the model parameters $\kappa$ and $\nu$. We note that the variance $\sigma^2$ is bounded for $\nu>0$ and $\kappa>1$, and that it is monotonically increasing in $\nu$, taking its maximum value $\sigma_{\max}=1/2\sqrt{3}$ as $\nu\to\infty$. In the large-$\nu$ limit, $\bar{\mu}\sim1/2$. We then evaluate the parameter regimes corresponding to the optimal migratory regime, \emph{i.e.}, solve for $\bar{\mu}(\kappa,\nu)=0.5$. When doing so, we find that there is an infinite set of parameters that satisfy these conditions and are parametrised by the curve:
\begin{equation}
\kappa^*(\nu)=-\frac{7\nu}{4}+\frac{1}{2}\sqrt{\frac{49\nu^2}{4}+8+20\nu}, \quad \nu>0.\label{limit A optimal}
\end{equation}
Expression~\eqref{limit A optimal} implies that the optimal value of $\kappa$ increases with $\nu$ ranging from $\kappa^*(0)=\sqrt{2}\approx1.414$ to $\kappa^*(\infty)\sim10/7\approx1.427$. Hence, the range of optimal $\kappa^*$ is narrow. Yet, particularly for large values of $\nu$, $\partial_\kappa\bar{\mu}$ is small, suggesting that deviation of $\kappa$ from the optimal value $\kappa^*$ does not significantly impact the mixing between the two. Hence, large levels of heterogeneity in the diffusivities of the two populations do not disrupt the co-localisation between the two populations, given that the spreading $\sigma$ remains finite and instead buffers deviations in the chemotactic sensitivities between the two populations. In contrast, the large-$\kappa$ regime distrupts the co-migration of of the two populations as both $\bar{\mu}$ and $\sigma^2$ decay as powers of $\kappa^{-1}$; specifically, $\bar{\mu}\sim(1+\nu)(2+3\nu)/2\kappa^2$ and $\sigma\sim \sqrt{P(\nu)}/2\kappa^2$, where $P(\nu)=\nu(\nu+1)(3\nu+2)(7\nu+4)$.

\subsection{Sublimit B: diffusive interface and short-range signalling}
We can use~\eqref{explicit solution large chi limit} to directly write $\eta$ as a function of the consumer cell density $\rho$ (or $S$ since the two are equivalent in this limit)
\begin{equation}
    \eta \sim K_\eta\left[(1-\rho)^{\kappa-1}\rho\right]^{1/\nu}.\label{eq:sensor distribution canosa}
\end{equation}
In this regime, unlike in regime A, the distribution of sensor cells relative to consumer cells directly follows a Beta distribution, characterised by the shape parameters $\beta=(\kappa-1)/\nu+1>1$ and $\alpha=1/\nu+1>1$. Therefore, the colocalisation metrics $\bar{\mu}$ and $\sigma^2$~\eqref{eq: definition colocalisation metrics} correspond respectively to mean and variance of the beta distribution $\text{\emph{Beta}}(\alpha,\beta)$:
\begin{equation}
\bar{\mu}=\frac{\alpha}{\alpha+\beta}=\frac{1+\nu}{\kappa+2\nu},\quad \sigma^2=\frac{\alpha\beta}{(\alpha+\beta)^2(\alpha+\beta+1)}=\frac{\nu(1+\nu)(\nu+\kappa-1)}{(\kappa+2\nu)^2(\kappa+3\nu)}.\label{canosa moments}%
\end{equation}
We can compute the range of parameters yielding the transition to sensor-led migration, \emph{i.e.} $\bar{\mu}(\kappa,\nu)=1/2$. We find that this requires $\kappa^*=2$, independently of the value of $\nu$. This is because the strength of the signal and the density of consumer cells $\rho$ decay at the same rate in front of the migrating front, so that effective chemotaxis of the sensor population is more strongly dependent on proximity to consumer cells. Moreover, Eq.~\eqref{canosa moments} reveals that co-migration is disrupted in the large-$\kappa$ limit as $\bar{\mu},\sigma\rightarrow 0$; specifically, $\mu\sim1/\kappa$ and $\sigma\sim\sqrt{\nu(\nu+1)}/\kappa$. This indicates complete demixing of the two populations, with the sensor population accumulating at the leading edge. Compared to the large-$\kappa$ limit in regime A, we find that $\bar{\mu}$ and $\sigma$ decay with $\kappa^{-1}$, \emph{i.e.}, less fast. Considering the large-$\nu$ limit instead, we recover the exact same limiting behaviour as for regime A, as $\bar{\mu}\to1/2$ and $\sigma^2\to1/2\sqrt{3}$.

\subsection{Sublimit C: sharp interface and long-range signalling precursor}
\label{sec: mixing sublimit C}
In this regime, in order to observe changes in $\rho$, we have to rescale the variables within the inner region and find that
\begin{equation}
    \eta\sim\begin{cases}
    e^{QZ}, &\quad Z\ll1\\
    e^{QZ-\kappa Z^2/2\nu}, &\quad Z\gg1.
    \end{cases}
\end{equation}
We find distinct regimes depending on the size of 
\begin{equation}
Q:=\frac{(\kappa-1)}{\nu}\sqrt{\frac{\chi_\rho}{D_\rho}},
\end{equation}
where we recall that $\sqrt{D_\rho/\chi_\rho}\ll1$.
Physically, if $Q\geq\mathcal{O}(1)$, the sensor population decay rate at the rear of the wave is comparable, or smaller, than the thickness of the interface. Therefore, the structure of $\rho$ across the interface is key to quantitatively estimating the mixing between the two populations. In contrast, if $Q\leq\mathcal{O}(1)$, the concentration of sensor cells is approximately constant across the interface. This can be achieved in two different ways: either $\kappa\sim 1$ or $\nu\gg\mathcal{O}(1)$, which correspond, respectively, to very small heterogeneity in the response of the two cells to the signal and very large heterogeneity in cell random motility. 

For the large-$Q$ regime, analytical progress can be made. In this regime, the mass of sensor cells that accumulate at the interface is infinitesimal. Hence, we can take  $\rho$ to be a Heaviside function, and $\eta$ will be distributed according to a Bernoulli random variable. Specifically, $\eta(\rho)\sim\text{Bernoulli}(p)$ where the parameter $p$ is equivalent to the fraction of sensor cells at the rear of the invading front:
\begin{equation}
    p=\frac{1}{1+I_+/I_-}, \qquad I_-=\int_{-\infty}^0 \exp\left[\frac{(\kappa-1)}{D_\eta}z\right]dx=\frac{D_\eta}{\kappa-1}, \qquad I_+=\int_{0}^{\infty}    \exp\left[-\dfrac{\kappa(z-z_\eta)^2}{2\chi_\rho D_\eta}+\dfrac{\kappa z_\eta^2}{2\chi_\rho D_\eta}\right]dz.
\end{equation}
This is also equivalent to the mean of the distribution, and therefore, our mixing metric
\begin{equation}
    \mu=p=\frac{1}{1+(\kappa-1)\left(\dfrac{eD^\kappa_\eta}{\chi^\kappa_\rho}\right)^{\frac{\chi_\rho}{D_\eta}}\Gamma\left(\dfrac{\kappa\chi_\rho}{D_\eta}+1,\dfrac{\chi_\rho}{D_\eta}\right)},\label{mixingC}
\end{equation}
where $\Gamma$ indicates the \emph{upper incomplete gamma} function. Expression~\eqref{mixingC} reveals that mixing of the two populations is primarily influenced by two key non-dimensional parameters: the relative chemotactic sensitivity of the two populations $\kappa$ and the rescaled diffusion coefficient $D_\eta/\chi_\rho$. As shown in Figure 3 of the main text, we find that the relative sensitivity to the chemoattractant $\kappa$ that corresponds to $\mu=0.5$ depends significantly on the sensor cells' diffusivity $D_\eta$. Taking the large-$D_\eta$ limit of~\eqref{mixingC}, we find that $\mu\sim\kappa^{-1}$. In contrast to regimes A-B, even for large values of the sensor motility $D_\eta/\chi_\rho\gg1$, the level of mixing between the two populations remains sensitive to the parameter $\kappa$, with the critical value of $\kappa_{cr}\to2$. Considering the large-$\kappa$ limit instead, we find that~\eqref{mixingC} decays to zero faster than exponentially: $\log\mu\sim -\chi_\rho\kappa/D_\rho\log\kappa$.

\subsection{Sublimit D: sharp interface and short-range signalling precursor}
In this regime, the analysis is similar to that in~\Cref{sec: mixing sublimit C}. The main difference is that, due to the different scaling of the interfacial thickness, the parameter $Q$ controlling the decay rate of the sensor cells is now defined as:
\begin{equation}
    Q:=\frac{(\kappa-1)}{\nu\sqrt{D_\rho}},
\end{equation}
and hence does not depend on $\chi_\rho$. As before, in the limit of $Q\ll\mathcal{O}(1)$ -- \emph{i.e.,} in the limit in which the decay length of the sensor cell distribution is larger than the thickness of the interface, the analysis of mixing simplifies, and the sensor population distributions follow a Bernoulli random variable with success probability $p$ corresponding to the fraction of cells located at the back of the wave:
\begin{equation}
    p=\frac{1}{1+I_+/I_-}, \qquad I_-=\frac{D_\eta}{\kappa-1}, \qquad I_+=\int_0^\infty\exp\left[-\dfrac{z-\kappa(1-e^{-z})}{D_\eta}\right]dz.
\end{equation}
By changing the integration variable, we can relate $I_+$ to well-known functions and obtain an expression for $p$; specifically
\begin{equation}
    \mu=p=\frac{1}{1+\dfrac{(\kappa-1) }{D_\eta}\left(\dfrac{e^{\kappa}D_\eta}{\kappa}\right)^{1/D_\eta}\gamma\left(\dfrac{1}{D_\eta},\dfrac{\kappa}{D_\eta}\right)}, \label{eq:mixing D}
\end{equation}
where $\gamma$ indicates the \emph{lower incomplete gamma} function. As shown, we find that the choice of the parameters that produce maximum mixing is now dictated both by the relative sensitivity of the population to the chemoattractant $\kappa$ and the diffusivity of the sensor population $D_\eta$. We find that the limiting behaviour of Eq.~\eqref{eq:mixing D} in the large-$D_\eta$ limit is similar to that of~\eqref{mixingC}. In the large-$\kappa$ limit, we find that $\mu$ decays exponentially with an algebraic correction that, depending on the strength of the sensor cell diffusion $D_\eta$, can either slow down (if $D_\eta<1$) or enhance (if $D_\eta>1$) the decay of $\mu$; specifically, $\mu\sim (\kappa/D_\eta)^{1/D_\eta-1}e^{-\kappa/D_\eta}$. While the reduction in the mixing with increased asymmetry in cell responses to the chemoattractant is faster than in the case of diffusive consumer fronts (regimes A and B), it is less than in regime C. This is due to the faster decay of the chemotactic signal ahead of the consumer front.

\end{document}